\documentclass[%
    aps,
    prb,
    reprint,
    superscriptaddress
    amsfonts,
    amsmath,
    amssymb,
    eprint,
    showkeys,
    longbibliography, 
    twocolumn,preprintnumbers,unsortedaddress
    ]{revtex4-2}

\usepackage{graphicx}
\usepackage{dcolumn}
\usepackage{amsmath,amsbsy,amssymb,scalefnt}
\usepackage{float} 
\usepackage{lipsum} 
\usepackage{verbatim}
\usepackage{bm}
\usepackage{epstopdf}
\usepackage{amsfonts}
\usepackage{textcomp}
\usepackage{soul}
\usepackage{mathtools}
\usepackage{hyperref}
\usepackage{comment}
\usepackage{fancyhdr}
\usepackage{mathtools}
\usepackage[normalem]{ulem}
\usepackage[x11names]{xcolor}
\hypersetup{
 pdfnewwindow=true, colorlinks=true,
 linkcolor=blue, anchorcolor=blue,
 citecolor=blue, filecolor=blue,
 menucolor=blue, urlcolor=blue
}

\usepackage{stackengine}

\usepackage{subcaption}
\usepackage{ragged2e}
\DeclareCaptionJustification{reallyjustified}{\justifying}
\graphicspath{{./Figures/}}

\newcommand{\leo}[1]{\textcolor{blue}{#1}} %

\begin{document}
\excludecomment{taglio}
\excludecomment{notainterna}
\title{Tuning Andreev reflection and conductance in   proximitized  nanowires\\ through the spin-orbit field direction}

\author{Leonardo Musca}

\author{Fabrizio Dolcini}
\email{fabrizio.dolcini@polito.it}

\affiliation{Dipartimento di Scienza Applicata e Tecnologia del Politecnico di Torino, I-10129 Torino, Italy}

\date{\today}

\begin{abstract}
We investigate the Andreev reflection (AR) and the nonlinear conductance in a normal-superconductor junction realized with  a proximitized semiconductor nanowire  exposed to a magnetic field along its axis. We show that, while at zero energy   the AR coefficient is determined by the topological phase of the superconducting side of the junction, at finite energy the AR is strongly affected by  the directions of the Rashba spin-orbit  field    characterizing the normal and the proximitized sides. The tunability  of the AR and the nonlinear conductance as a function of the spin-orbit misalignment angle  is particularly pronounced when the proximitized side is in the  topologically trivial phase. The spin-orbit field direction   is revealed to be  an efficient knob to  control the electron transport through the hybrid junction, realizing an electrically widely tunable device. 
The implementation in realistic setups is discussed.
\end{abstract}

\maketitle

\section{Introduction}
Semiconductor Nanowires (NWs) with strong spin--orbit coupling have emerged as a versatile platform to explore fundamental quantum transport phenomena and to engineer  novel quantum devices. This is due to two important factors. First, owing to their quasi-one-dimensional geometry and high crystalline quality~\cite{fan_2015},   NWs exhibit electron ballistic propagation \cite{yu_2006,javey_2013,schaepers_2016,kouwenhoven_natnano_2018}, allowing for the observation of phase coherent interference effects~\cite{kouwenhoven_2009,schaepers_2010,schaepers_2011,schaepers_2012,xu_2015,qi_2017}   and quantized conductance~\cite{kouwenhoven_2013,kouwenhoven_2016,kouwenhoven_2017}. Second, the Rashba spin--orbit (SO) field originating from structural inversion asymmetry (SIA) is tunable via electrostatic gates in different geometries \cite{goldoni_2018,loss-prb-2018c,gao_2012,kouwenhoven-wimmer_2015,kouwenhoven-natcomm_2017,sasaki_2013,sasaki_2017,sasaki_2021,guo_2021,moon_2006,wernersson_2008,wernersson_2010,deshmukh_2011,micolich_2015,nygard_2016,strambini_2018,burke_2019}, making NWs highly adaptable  and particularly attractive quantum systems for studying spin-dependent transport at the nanoscale~\cite{loss_2008,vanweert_2009,vandenberg_2013,pribiag_2013}.

Because of these remarkable features, NWs have found applications in multiple fields. In nanoelectronics, they serve as building blocks for field-effect transistors, with excellent electrostatic control and scalability~\cite{beltram_2012,shimoida2013,javey_2013,takase2017}. Also, their one-dimensional geometry and tunable carrier density   make them promising candidates for thermoelectric devices, where quantum confinement can enhance thermopower and energy conversion efficiency~\cite{riel_2013,sergej_2020,elsachat_2021,taddei_2024,aslani_2026}. 
Furthermore, by
covering a  NW with a superconductor, the interplay between the  superconducting pairing induced by proximity effect, the Rashba SO field and  an externally applied magnetic field  realizes a topological superconductor~\cite{dassrma_prl_2010,vonoppen_prl_2010}. The  Majorana  modes predicted to emerge at the NW ends could have quite relevant applications in topological quantum computation~\cite{alicea_2012,alicea_prx_2016,aguado_2017,dassarma_prb_2018}.

Although an unambiguous experimental evidence of the  exotic Majorana quasiparticles is still a debated issue~\cite{loss_prb_2017,loss_prb_2018a,loss_prb_2018b,dassarma_prb_2018,prada_2020}, the extensive effort devoted in their search has also boosted the  investigation of {\it hybrid} NW devices, where only some segments of the NW are proximitized by a  superconductor(S), while  other segments remain unproximitized and are dubbed   normal(N)~\cite{heiblum_2012,kouwenhoven_science_2012,furdyna_natphys_2012,xu_2012,nilsson_2012,marcus_2016,kouwenhoven_nanolett_2017,kouwenhoven_natnano_2018}. Various interesting effects emerge in such hybrid NW junctions. 
For instance, it has been theoretically predicted~\cite{levy-yeyati_2017,pothier_2021,houzet-meyer_2024} and experimentally proven~\cite{pothier_2019,hays_2020,hays_2021,pita-vidal_2023,kouwenhoven_2023,pita-vidal_2024,pita-vidal_2025,fatemi_prappl_2025,fatemi_prl_2025} that S-N-S junctions based on NWs can be harnessed to realize Andreev spin qubits, which can be readout and manipulated via microwave radiation. Also, such  junctions can exhibit anomalous Josephson effect~\cite{lucignano_2015,houzet-meyer_2016} and, when voltage-biased, the resulting  dc current  can reveal information about the  topological  transition occurring in the S-segment~\cite{aguado_njp_2013}.
Furthermore, recent studies on NW-based N-S-N junctions  have revealed the competition between crossed  AR and elastic cotunneling~\cite{kouwenhoven_prx_2023,loss-ando_2025}.

\begin{figure}[h]
\includegraphics[width=\linewidth]
{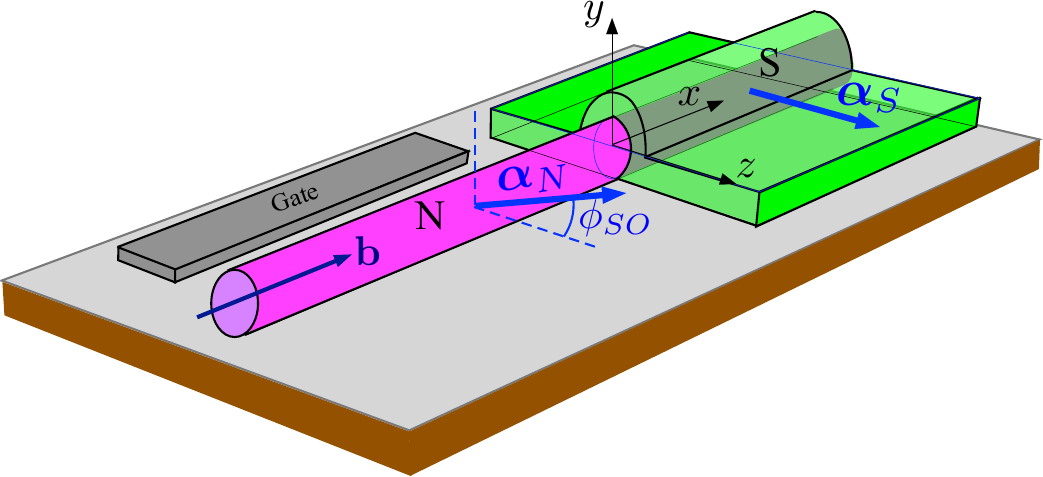}
\caption{\label{Fig1-setup} Scheme of   a hybrid N-S junction based on a semiconductor NW with SO coupling. The S-side is proximitized by a superconducting film (green area). The SO field vectors $\boldsymbol{\alpha}_N$ and $\boldsymbol{\alpha}_S$ lie in the $y$-$z$ plane and point in two different directions, differing by a misalignment angle $\phi_{SO}$. A magnetic field $\mathbf{b}$ directed along the NW axis is externally applied.  }
\end{figure}

At any N-S interface in a NW, spatial inhomogeneities in the SO field naturally arise. Indeed the SIA in the uncovered N-side differs from the one in the covered proximitized S-side, owing to changes in electrostatic screening, dielectric environment, to the geometry of gates coupled to the N-side, or to the way the superconducting film is deposited  onto the S-side. Such inhomogeneities are expected to affect the AR process, since an incident electron from the N-side is  reflected as a hole in a way that is sensitive to spin structure, which in turn depends on the SO field. Two types of inhomogeneity can arise across the N-S interface, namely  in the magnitude  and in the direction of the SO-field. The former type of inhomogeneity is expected to affect AR for trivial reasons: The  magnitude of the local SO field directly modifies the   spectrum of the related NW segment, giving rise to a potential step barrier at the interface. This effect already exists in normal NWs, where local inhomogeneities of the SO magnitude have been shown to affect the conductance and the spin density profile~\cite{sanchez_2006,sanchez_2007,sanchez_2008,sherman_prb_2013,tokatly_2017,rossi-dolcini_prb_2018,rossi-dolcini_prb_2020,rossi-dolcini_epjb_2020}. 
For this reason,   most experimental efforts have been devoted to enhancing the range of tunability of the SO magnitude, exploiting various gating techniques~\cite{gao_2012,kouwenhoven-wimmer_2015,kouwenhoven-natcomm_2017,sasaki_2013,sasaki_2017,sasaki_2021,guo_2021}.

By contrast, the latter type of SO-inhomogeneity --in the SO field direction-- is more subtle:  Although it leaves  the local spectrum of the NW segment unaltered,  it does  affect  the electron eigenfunctions in a nontrivial way. Notably, while  experimental advances have been made to control  the SO field direction  with suitable gating geometries, such as side  or wrap gates~\cite{moon_2006,wernersson_2008,wernersson_2010,deshmukh_2011,micolich_2015,nygard_2016,strambini_2018,burke_2019}, or with a
lateral deposition of the proximitizing superconducting film~\cite{kouwenhoven_prl_2019,kouwenhoven_prappl_2024},   theoretical studies  have analyzed this problem  only partially so far, and  mainly in two cases. On the one hand, in a normal NW, an antiparallel SO field configuration can realize a Dirac paradox configuration for electrons, which has been  predicted to strongly reduce the conductance and to dramatically enhance the thermoelectrical properties of the NW~\cite{gogin2022a,gogin2022b,aslani_2026}. On the other hand, in the presence of a superconducting pairing present in the entire NW, a SO direction inhomogeneity   leads to trivial  fermionic bound states that, at anti-parallel configuration, are degenerate with Majorana quasiparticles~\cite{loss-epjb-2015}.
However,  when it comes to hybrid N-S junctions based on NWs, the effects of the SO field direction inhomogeneities on the AR have not been investigated in details yet. Because the AR process is the fundamental mechanism governing the electronic behavior of hybrid quantum devices, understanding such effect is essential for interpreting experiments  and for optimizing the hybrid device performance in applications, ranging from Majorana platforms to Andreev  spin qubits.

This paper aims to address this problem. We analyze a N-S junction based on a single-channel NW, exposed to a Zeeman field $\mathbf{b}$ applied along the NW axis, as depicted in Fig.\ref{Fig1-setup}. We show that, while at zero energy   the AR coefficient is purely determined by the topological phase of the superconducting side, at finite energy the AR and conductance are strongly determined by the SO field misalignment in the two sides of the interface. In particular, the trivial regime seems to be particularly suitable to yield a widely tunable AR and conductance at finite energy. These results indicate that  the spin-orbit field  direction can be harnessed as a knob to  realize an electrically controllable hybrid device.

The paper is organized as follows. In Sec.\ref{sec-2} we present the model for a inhomogeneous proximitized NW. In Sec.\ref{sec-3}, after summarizing the main ingredients of the Scattering matrix method applied to the NW case, we present  our results about the  AR coefficient $R_{he}$ and the nonlinear conductance $G$ of the  N-S junction, showing their dependence on the relative SO angle, for various values of the Zeeman field, exploring both the topological and the trivial phase of the S-side. Then, in Sec.\ref{sec-4}, we discuss our results, provide an interpretation for the tunability of $R_{he}$ and  $G$, and propose implementation schemes in realistic setups. Finally, we draw our conclusions in Sec.\ref{sec-5}.\\

\section{Model for a proximitized NW with inhomogeneous spin-orbit field}
\label{sec-2}
We consider a single-channel NW deposited on a substrate and denote by $x$ the direction of its axis, by $z$ the other direction lying in the substrate plane, and by $y$ the direction perpendicular to the substrate, as illustrated in Fig.\ref{Fig1-setup}. The NW Hamiltonian consists of four terms,
\begin{equation}\label{H-NW-sum}
\mathcal{H} = \mathcal{H}_{kin}
+ \mathcal{H}_{SO} +\mathcal{H}_{Z} +\mathcal{H}_{SC}\quad,
\end{equation}
where
\begin{equation}\label{H-NW-kin-term}
\mathcal{H}_{kin} = 
\int \! dx \left(\Psi^\dagger_{\uparrow}(x),\Psi^\dagger_{\downarrow}(x)\right) \left( \frac{\hat{p}^2}{2m^*} -\mu(x)\! \right)  \sigma_0 \! 
    \begin{pmatrix}
        \Psi_{\uparrow}(x)\\
        \Psi_{\downarrow}(x)
    \end{pmatrix}  
\end{equation}
is the kinetic term in the effective mass $m^*$ approximation. Here, $\Psi_{\sigma}(x)$ denotes the electron field operator for spin $\sigma=\uparrow,\downarrow$, while $\hat{p}=-i \hbar \partial_x$ is the momentum operator, and $\mu(x)$ denotes  the chemical potential profile. The second term in Eq.(\ref{H-NW-sum})  is
\begin{equation}\label{H-NW-SO-term}
\mathcal{H}_{SO} =- \frac{1}{2\hbar} \!
\int dx \left(\Psi^\dagger_{\uparrow}(x),\Psi^\dagger_{\downarrow}(x)\right) \left\{  \boldsymbol{\alpha} (x),\hat{p}\right\}   \cdot \boldsymbol{\sigma}  
    \begin{pmatrix}
        \Psi_{\uparrow}(x)\\
        \Psi_{\downarrow}(x)
    \end{pmatrix}
\end{equation}
and describes the inhomogeneous Rashba SO coupling, characterized by a space-dependent SO vector field $\boldsymbol{\alpha} (x)$. The anticommutator  $\{\cdot\, , \cdot\}$  term   is necessary because $\boldsymbol{\alpha}(x)$ does not commute with the momentum $\hat{p}$. Moreover, 
\begin{equation}\label{H-NW-Z-term}
\mathcal{H}_{Z} = -\int dx  \left(\Psi^\dagger_{\uparrow}(x),\Psi^\dagger_{\downarrow}(x)\right) \boldsymbol{b}   \cdot \boldsymbol{\sigma}  
    \begin{pmatrix}
        \Psi_{\uparrow}(x)\\
        \Psi_{\downarrow}(x)
    \end{pmatrix} 
    \end{equation}
denotes the Zeeman coupling with the applied magnetic field, whereas
\begin{equation}
\mathcal{H}_{SC} = \int dx \Big[\Delta(x) \Psi^\dagger_{\uparrow}(x)\Psi^\dagger_{\downarrow}(x) + \Delta^*(x) \Psi_{\downarrow}(x) \Psi_{\uparrow}(x) \Big]\label{H-NW-SC-term}
\end{equation}
is the superconducting pairing potential induced by proximity effect,
which is assumed to vanish  in the unproximitized N-side on the left of the interface, located at $x_0$, while it is present in the NW proximitized side S on the right of the interface\begin{equation}
\Delta(x)=\left\{ \begin{array}{lcl} 0 & & x<x_0 \\
\Delta_0 e^{i \varphi}& & x>x_0 \end{array} \right.\quad.
\end{equation}

By introducing the  four components   Nambu spinor field operator
\begin{equation}\label{Nambu-spinor-real-space}
\Psi(x) =\left(
\begin{array}{c}
    \Psi_{\uparrow}(x) \\
    \Psi_{\downarrow}(x) \\[2pt]
    \Psi^\dagger_{\downarrow}(x) \\
    -\Psi^\dagger_{\uparrow}(x)
\end{array}
\right)\quad,
\end{equation}
the  proximitized NW Hamiltonian Eq.(\ref{H-NW-sum}) can be rewritten, up to an additive constant, in the Bogolubov -- de Gennes (BdG) form~\cite{bogoljubov_1958,degennes_book,ketterson_book}
\begin{equation}\label{H-NW-inhomo-1}
    \mathcal{H} = \frac{1}{2}\int dx\,\Psi^\dagger(x) \, H_{BdG} (x) \Psi(x) \quad,  
\end{equation}
where the BdG  Hamiltonian
\begin{equation}\label{HBdG-structure-v1}
H_{BdG}(x)=\begin{pmatrix}
         h_e(x)   & \Delta(x) \sigma_0
     \\
     \Delta^*(x) \sigma_0 &   h_h(x) 
     \end{pmatrix}
\end{equation}
is characterized by a  $2 \times 2$ electron-block $h_{e}(x)$ and a $2 \times 2$ hole-block   $h_{h}(x)$, mutually related by
\begin{equation}\label{electron-to-hole-relation}
h_{h}(x) = -\sigma_y h^*_{e}(x)\sigma_y \quad,
\end{equation}
and explicitly given by 
\begin{align}
h_{e}(x) 
&=\label{h_e_matrix-NW}
\left(-\frac{1}{2m^*} \frac{\partial^2}{\partial x^2}-\mu(x) \right)\sigma_0  - \mathbf{b}_{SO}(x)  \cdot \boldsymbol{\sigma}   - \mathbf{b}  \cdot \boldsymbol{\sigma}
  \\[8pt]
h_{h}(x) 
&=\label{h_h_matrix-NW}
-\left(-\frac{1}{2m^*} \frac{\partial^2}{\partial x^2}-\mu(x) \right) \sigma_0  + \mathbf{b}_{SO}(x)  \cdot \boldsymbol{\sigma} - \mathbf{b}  \cdot \boldsymbol{\sigma}
\end{align}
respectively. 
In Eqs.(\ref{h_e_matrix-NW})-(\ref{h_h_matrix-NW}),  
\begin{eqnarray}
\hat{\mathbf{b}}_{SO}(x) &=& \frac{1}{2\hbar} \Big\{ \boldsymbol{\alpha}(x),\hat{p}\Big\} = -\frac{i}{2} \Big\{ \boldsymbol{\alpha}(x),\frac{\partial}{\partial x}\Big\} = \nonumber \\
&=& -i \left(\boldsymbol{\alpha}(x) \frac{\partial}{\partial x}+\frac{1}{2}(\partial_x \boldsymbol{\alpha}(x) ) \right)\label{bSO(x)-def}
\end{eqnarray}
describes the SO field operator. Note that, differently from the actual Zeeman field $\mathbf{b}$, the  SO field $\hat{\mathbf{b}}_{SO}(x)$ is  complex and, in particular,  purely imaginary, i.e. $\hat{\mathbf{b}}^*_{SO}(x)=-\hat{\mathbf{b}}_{SO}(x)$.\\

Then, the BdG Hamiltonian (\ref{HBdG-structure-v1}) entails the BdG Equations,
\begin{equation}\label{BdG-eq}
 H_{BdG}(x)\Phi(x)=E \,\Phi(x) \quad E \ge 0\quad,
\end{equation}
where 
\begin{equation}\label{BdG-Phi}
\Phi(x)=\left(\begin{array}{c}
 \begin{array}{c}\mathbf{u}(x) \end{array}   \\
  \begin{array}{c} \mathbf{v}(x) \end{array}  \end{array}\right) =\left(\begin{array}{c}
 {u}_\uparrow(x)   \\  {u}_\downarrow(x)  \\
 {v}_\downarrow(x)   \\  {v}_\uparrow(x)  \end{array}\right) 
\end{equation}
is the  $4 \times 1$  eigenfunction, whose  upper $2 \times 1$ spinor $\mathbf{u}(x)$ and  lower $2 \times 1$ spinor $\mathbf{v}(x)$ refer to the electron and hole eigenfunction components, respectively.

\subsection{Magnetic field along the NW axis and piecewise inhomogeneous spin-orbit coupling}
While the inhomogeneous model in Eq.(\ref{H-NW-sum}) is actually quite general, we shall henceforth make specific assumptions that suit for the N-S setup we aim to consider here. First, we shall consider the case where the magnetic field is applied along the NW axis $x$
\begin{equation}
\mathbf{b}=(b_x,0,0) \quad.
\end{equation} 
Second, recalling that the Rashba SO field  is $\boldsymbol{\alpha} \propto \mathbf{k} \times   \mathbf{E}$, where $\mathbf{k}=(k_x,0,0)$ is the propagation wavevector along the single-channel NW axis and $\mathbf{E}$ is the SIA electric field, one deduces that the SO field $\boldsymbol{\alpha}(x)$ lies entirely in the $(y,z)$ plane perpendicular to the electron propagation direction $x$ (see Fig.\ref{Fig1-setup}). This implies that the magnetic field and the SO coupling are mutually orthogonal.
Third, we assume that   the SO field $\boldsymbol{\alpha}(x)$ can be fairly approximated by a   piecewise constant profile   
\begin{equation} 
\boldsymbol{\alpha}(x) = \left\{ \begin{array}{lcl}   \boldsymbol{\alpha}_N  & \mbox{for} &   x < x_0 \\ & & \\
\boldsymbol{\alpha}_S & \mbox{for} & x> x_0  \end{array}
\right. \quad, \label{alpha-piecewise}
\end{equation}  
where   $\boldsymbol{\alpha}_N$ and $\boldsymbol{\alpha}_S$ denote the SO fields in the N-side and in the proximitized S-side of the interface. The fields $\boldsymbol{\alpha}_N$ and $\boldsymbol{\alpha}_S$ can  differ by
magnitude and orientation. Without loss in generality, we shall take
\begin{eqnarray}\label{alphaN-N-vec-def}
\boldsymbol{\alpha}_N   &=& \alpha_N  \left(0 , \sin \phi_{SO}, \cos \phi_{SO}\right)\\ & & \nonumber \\
\boldsymbol{\alpha}_S  & = &\left(0 , 0, \alpha_S \right)  \label{alphaN-S-vec-def} \quad,
\end{eqnarray}
with $\alpha_N>0,\alpha_S>0$, while $\phi_{SO}$ describes the misalignament angle between the two SO field directions.

A piece-wise constant profile is also assumed for the chemical potential $\mu(x)$, with $\mu_N$ and $\mu_S$ denoting its values on the N and S-sides, respectively. 
As a consequence, the solution $\Phi(x)$ of the BdG equations (\ref{BdG-eq}) for the entire N-S system is expressed as a piecewise function
\begin{equation}
\Phi (x) = \left\{
    \begin{array}{lcl}
        \Phi_N(x) & \mbox{if} & x < x_0  \\\\
         \Phi_S(x) & \mbox{if} & x>x_0
    \end{array}\right.\quad,
\end{equation}
where   $\Phi_N$ (N-side) and $\Phi_S$ (S-side) fulfill  boundary conditions. Explicitly, while the wavefunction is continuous,
\begin{equation}\label{bc-fun}
     \displaystyle   \Phi_N(x_0) = \Phi_S(x_0)  \quad,
\end{equation}
its derivative exhibits a discontinuity at the interface location $x_0$, related to the   inhomogeneity of the SO field~(\ref{alpha-piecewise})
\begin{eqnarray}
    \displaystyle \partial_x \Phi_N(x_0) -i \frac{m^*}{\hbar^2} \Big(\tau_0 \otimes \big( \boldsymbol{\alpha}_N \cdot\boldsymbol{\sigma}\big)\Big) \,\Phi_N(x_0)  = \nonumber \\
    \partial_x \Phi_S(x_0) -i \frac{m^*}{\hbar^2} \Big(\tau_0 \otimes \big( \boldsymbol{\alpha}_S \cdot\boldsymbol{\sigma}\big)\Big) \,\Phi_S(x_0)\label{bc-der} \quad.
\end{eqnarray}
Here, $\tau_0$ denotes the identity in Nambu space.  \\

\subsection{Excitation spectra and energy regimes in the two sides of the junction}
Here, we briefly recall  the main aspects characterizing the excitation spectra and the eigenstates   in the bulk of each junction side. 

As far as the bulk spectrum is concerned,  it consists 
of the   positive energy eigenvalues $E \ge 0$ 
 obtained by plugging a plane-wave solution $\Phi_k(x)=\mathsf{w}(k) e^{i k x}$ into the BdG equations (\ref{BdG-eq}), where $\mathsf{w}_E$ denotes a $4\times 1$ spinor.
Specifically, on the N-side, the spectrum is 
\begin{equation}\label{spectrum-N-side}
E_{N,\pm}(k)= \left|\varepsilon_0(k)-\mu_N \pm \sqrt{4   \varepsilon_0(k)\, E_{SO,N}+ E_Z^2}  \right|\quad,
\end{equation}
whereas  on the S-side one finds~\cite{vonoppen_prl_2010,dassrma_prl_2010}
\begin{eqnarray}
\lefteqn{E_{S,\pm}(k) =  \left[ (\varepsilon_0(k)-\mu_S)^2+4 \varepsilon_0(k) \, E_{SO,S}    + E_Z^2    \right.  } & &   \label{spectrum-S-side} \\
& & \left. \hspace{-0.5cm} + \Delta_0^2 \pm 2\sqrt{(\varepsilon_0(k)-\mu_S)^2 (4 \varepsilon_0(k) \, E_{SO,S}  + E_Z^2) +E_Z^2\Delta_0^2}\right]^{1/2}  .\nonumber   
\end{eqnarray}
Here, $\varepsilon_0(k)=\hbar^2 k^2/2m^*$, while the index   $b=\pm$ denotes the two bands. Furthermore, 
\begin{equation}
E_{Z} = |b_x|
\end{equation}
is the Zeeman energy related to the applied magnetic field, while
\begin{equation}\label{ESOj-def}
E_{SO,j}= \frac{m^* |\boldsymbol\alpha_j|^2}{2 \hbar^2} \hspace{1cm} j=N,S
\end{equation}
denotes the SO energy on the two sides of the hybrid junction. 

We recall that the spectrum (\ref{spectrum-N-side}) on the N-side can exhibit  two possible regimes~\cite{streda_2003,houzet-meyer_2016,rossi-dolcini_prb_2020}, namely  the Zeeman-dominated regime ($2E_{SO,N}<E_Z$), and the Rashba-dominated regime ($2E_{SO,N}>E_Z$), where the lower excitation   band ($b=-$) exhibits a mexican-hat shaped behavior as a function of $k$.  In contrast, the spectrum (\ref{spectrum-S-side}) of the S-side is characterized by a    superconducting gap $\Delta_S=|E_Z-E_Z^c|$ at $k=0$, closing at the critical value \begin{equation}
\label{EZc} E_Z^c=\sqrt{\Delta_0^2+\mu_{S}^2}\quad.
\end{equation}
The value $E_Z^c$ is known to separate the trivial phase ($E_Z< E_Z^c$) from the topological phase  ($E_Z> E_Z^c$) that hosts Majorana bound states at the S ends \cite{dassrma_prl_2010,vonoppen_prl_2010}.
Examples of the   excitations branches in the bulks of the N and S sides are depicted in panels (a) of Figs.\ref{Fig-2-Topological} and  \ref{Fig-3-Trivial}.

When it comes to the bulk eigenstates on each junction side,   as is customary in hybrid junctions involving superconductors, the excitation wavefunction  $\Phi_k(x)=\mathsf{w}(k) e^{i k x}$  can be also attributed a  quantum number $p=e/h$ identifying the predominant electron(e)-like   or  hole(h)-like  character of the excitation. Explicitly, denoting the spinor components for each band   $b=\pm$  as
\begin{equation}
    \mathsf{w}_\pm(k)=\begin{pmatrix}
        u_{\pm,\uparrow} \\ u_{\pm,\downarrow} \\
        v_{\pm,\downarrow} \\
        v_{\pm,\uparrow}  
    \end{pmatrix}\quad,
\end{equation}
the sign of quantity
\begin{equation}\label{charge character}
   \mathcal{Q}_\pm(k)=  |{u}_{\pm,\uparrow}(k)|^2 + |{u}_{\pm,\downarrow}(k)|^2  -  |{v}_{\pm,\uparrow}(k)|^2
    - |{v}_{\pm,\downarrow}(k)|^2 
\end{equation}
determines whether $\mathsf{w}_\pm(k)$ has an electron-like character $p=e$ (if $\mathcal{Q}_\pm>0$), or a hole-like character $p=h$ (if $\mathcal{Q}_\pm<0$). In the N-side, one has either $\mathcal{Q}_\pm=+1$ or $\mathcal{Q}_\pm=-1$, corresponding to purely electron or purely hole excitations.  In the Appendix~\ref{AppA}, the expressions of the resulting spinors $\mathsf{w}_{e/h, \pm}(k)$ are explicitly provided for both junction sides.

\section{Method and Results}
\label{sec-3}
In this section we shall present our results about the  AR and the nonlinear conductance of the 
NW-based N-S junction. 
We start by observing that, because  the SO energies (\ref{ESOj-def}) depend only on the {\it magnitude} $|\boldsymbol\alpha|$ of the SO field $\boldsymbol\alpha$, a difference in  $|\boldsymbol\alpha(x)|$  across the interface  directly affects the spectra   (\ref{spectrum-N-side}) and (\ref{spectrum-S-side}) on the two junction sides, giving rise to an effective potential step. As mentioned above, such inhomogeneities of the SO magnitude have already been studied  in normal NWs and, as expected, they can affect the transport properties and the spin density profile~\cite{sanchez_2006,sanchez_2007,sanchez_2008,sherman_prb_2013,tokatly_2017,rossi-dolcini_prb_2018,rossi-dolcini_prb_2020, rossi-dolcini_epjb_2020}. 
By contrast,   the eigenstates $\Phi_k(x)=\mathsf{w}(k) e^{i k x}$ also depend on
the {\it direction} of~$\boldsymbol\alpha$, through the spinor $\mathsf{w}(k)$. Thus, due to  the boundary conditions (\ref{bc-fun})-(\ref{bc-der}) of the wavefunctions,   transport properties turn out to depend non-trivially on the relative angle $\phi_{SO}$ characterizing the two directions in Eqs.(\ref{alphaN-N-vec-def}). 

The goal of our investigation is point out that such a dependence on $\phi_{SO}$ is strong, and that it can be harnessed to tune the   AR and  the nonlinear conductance of the NW. 
Here below, after describing the method adopted to  investigate the hybrid  junction and   defining the analyzed quantities,  our results are illustrated.

\subsection{Scattering Matrix approach}
\label{sec3a}
Our approach is based on solving the BdG equations (\ref{BdG-eq}) through the   Scattering matrix formalism for hydrid normal/superconducting systems~\cite{lambert_1991,BTK_1992,beenakker_1992,lambert_1993,lambert_1993-2,lambert_1996,datta-anantram_1996,lambert_1998}. To this purpose, we have  complementarily combined analytical approaches and numerical methods, whose main ingredients are   summarized here below,  where we also specify the adopted notation. The interested reader can find further technical details in the Appendix~\ref{AppB}.

At a given energy $E$, the wavefunction  $\Phi_N(x)$ and $\Phi_S(x)$ on each side $L=N,S$ of the junction can always be expressed as a superposition of planewave  $4 \times 1$ eigenstates of the form $\mathsf{w}_E\, e^{i k_E x}$. Each wave, weighted by a  related complex amplitude, can be classified according to various quantum numbers. Indeed, for each band $b=\pm$, various $k(E)$ wavevectors are obtained by solving the equation $E_{L,\pm}(k)=E$  [see Eqs.(\ref{spectrum-N-side}) and (\ref{spectrum-S-side})],   on each side $L=N,S$ of the interface. 
At    energy $E$, wavefunctions characterized by a real wavevector $k(E)$ identify the  propagating modes, which appear always in pairs $(k(E),-k(E))$ because Eqs.(\ref{spectrum-N-side}) and (\ref{spectrum-S-side}) are even in $k$. The sign of the group velocity $v=\hbar^{-1} \partial_k E_k$ identifies the propagation direction of the mode (rightwards or leftwards). Furthermore, at the energy $E$, one can also have   wavevectors $k$ with an imaginary part. Although such wavevectors  always come in complex conjugate pairs $(k^{}(E),k^*(E))$, only the ones that do not imply an exponential divergence of their wavefunction at $|x|\rightarrow \infty$  are   physically acceptable and are the   evanescent modes.
Finally, modes at energy $E$ can further be classified as  electron  or hole  modes  on the N-side,  and  electron-like  or  hole-like modes on the S-side. 

Henceforth, we shall label by $a$ the amplitudes of   propagating modes that are {\it incoming} towards the interface,   by $b$ the amplitudes of the modes that are {\it outgoing} from the interface, while the   letter $c$ will be devoted to the amplitude of evanescent modes.
Thus, each of these states is labelled by a quantum number set $(L,p,\lambda)$, where  $L=N,S$ labels the side of the junction, $p=e/h$ the electron-like or hole-like character, while $\lambda$ collects the band index $b=\pm$ and all  the other possible different $k$-values of that $(L,p)$-type.

Explicitly, on the N-side, the wavefunction at a given energy $E$ reads
\begin{eqnarray}
\lefteqn{\Phi_N (x)  = \sum_{p=e,h} \frac{1}{\sqrt{2 \pi \hbar}} \times } & & \label{Generic-wvf-N-side}  \\
 & &  \times     \left\{ \sum_{\lambda^a} \frac{a_{N,p,\lambda^a}}{\sqrt{|v_{N,p,\lambda^a}|}}\mathsf{w}_{N,p,\lambda^a}(k^a_{N,p,\lambda^a}) e^{i k^a_{N,p,\lambda^a} x} + \right. \nonumber \\
       & &  \hspace{0.3cm} +\sum_{\lambda^b}  \frac{b_{N,p,\lambda^b}}{\sqrt{  |v_{N,p,\lambda^b}|}}\mathsf{w}_{N,p,\lambda^b}(k^b_{N,p,\lambda^b}) e^{ik^b_{N,p,\lambda^b} x} + \nonumber \\
    & & \left. \hspace{0.3cm} +\sum_{\lambda^c}  \frac{c_{N,p,\lambda^c}}{\sqrt{|v_{N}|}}\mathsf{w}_{N,p,\lambda^c}(k^c_{N,p,\lambda^c}) e^{ik^c_{N,p,\lambda^c} x} \right\}\quad, \nonumber 
    \end{eqnarray}
where  the amplitudes  $a_{N,p,\lambda^a}$ and $b_{N,p,\lambda^b}$ characterize   right-moving modes incoming from the N-side towards the interface and left-moving modes outgoing from the interface into N, respectively, while  $c_{N,p,\lambda^c}$   labels the  amplitude of the evanescent modes decaying at $x\rightarrow -\infty$.  In Eq.(\ref{Generic-wvf-N-side}), $v_{N,p,\lambda^a}, v_{N,p,\lambda^b}$ denote the   group velocities of the propagating modes, with momenta $k^a_{N,p,\lambda^a}, k^b_{N,p,\lambda^b} \in \mathbb{R}$, while   for the evanescent modes  the momenta $k^c_{N,p,\lambda^c}$  exhibit a negative imaginary part, and the velocity $v_N = \sqrt{2(E_{SO,N}+|\mu_N|)/m^* }$  is formally introduced  just to have the $c$-amplitudes with the same units as the amplitudes $a$ and~$b$. 

An expression similar to Eq.(\ref{Generic-wvf-N-side}) holds in the S-side,   upon an appropriate relabelling ($N\rightarrow S$) in the  amplitude coefficients. In particular, $k^c_{S,p,\lambda^c}$ exhibits a positive imaginary part. For energies values below the superconducting  gap of the S-side, which in general depends on $\Delta_0, E_Z$ and $E_{SO,S}$, there are only evanescent modes. \\

The boundary conditions (\ref{bc-fun})-(\ref{bc-der}) enable one to express all the $b$ amplitudes of the   outgoing  propagating modes and all the $c$ amplitudes of the evanescent modes in terms of the $a$ amplitudes of the  incoming  propagating modes. Details about this derivation are given in Appendix~\ref{AppB}. Denoting by $\mathbf{b}_{L,p}=(b_{L,p,\lambda^b_1}, b_{L,p,\lambda^b_2}\ldots)^T$ and $\mathbf{a}_{L,p}=(b_{L,p,\lambda^a_1}, b_{L,p,\lambda^a_2}\ldots)^T$  the vectors of outgoing mode amplitudes and incoming mode amplitudes in the lead $L=N,S$, for particle nature $p=e,h$, they are connected through the Scattering matrix as follows
\begin{equation}\label{S-matrix}
    \begin{pmatrix}
        \mathbf{b}_{N,e} \\ \mathbf{b}_{N,h} \\ \mathbf{b}_{S,e} \\ \mathbf{b}_{S,h}
    \end{pmatrix} =\left(\begin{array}{cc|cc}
\mathsf{r}_{ee} & \mathsf{r}_{eh}      &\mathsf{t}^\prime_{ee} & \mathsf{t}^\prime_{eh} \\
  \mathsf{r}_{he} & \mathsf{r}_{hh} &  \mathsf{t}^\prime_{he} & \mathsf{t}^\prime_{hh} \\  \hline
\mathsf{t}_{ee} & \mathsf{t}_{eh} &  \mathsf{r}^\prime_{ee} & \mathsf{r}^\prime_{eh} \\  
\mathsf{t}_{he} & \mathsf{t}_{hh} &  \mathsf{r}^\prime_{he} & \mathsf{r}^\prime_{hh}
    \end{array} \right)\begin{pmatrix}
        \mathbf{a}_{N,e} \\ \mathbf{a}_{N,h} \\ \mathbf{a}_{S,e} \\ \mathbf{a}_{S,h}
    \end{pmatrix}\quad.
\end{equation}
Here, the entries $(\mathsf{r}_{he})_{\lambda\mu}$ of the block matrix $\mathsf{r}_{he}$  describes the AR process  where an electron state $\lambda$ injected from the N-side  gets reflected into a hole state $\mu$ available in~N. Similarly, the  $\mathsf{r}_{eh}$ block also describes an AR process, where hole and electron roles are interchanged though. The $\mathsf{r}_{ee}$ and $\mathsf{r}_{hh}$ blocks account for  the normal electron-electron and hole-hole reflections in N, while $\mathsf{t}$ and $\mathsf{t}^\prime$ blocks describe transmission from N to S and viceversa, respectively. For energies $E$ below the superconducting gap of the S-side, there exist no propagating modes in S, the $\mathbf{a}_{S,p}$ and $\mathbf{b}_{S,p}$ amplitude vectors are void, and the Scattering matrix only contains the left upper block of reflections.

In particular, the total $e \rightarrow h$ AR coefficient is defined as \cite{lambert_1991,BTK_1992,beenakker_1992,lambert_1993,lambert_1993-2,lambert_1996,datta-anantram_1996,lambert_1998}
\begin{equation}
R_{he}(E) = {\rm Tr} (\mathsf{r}^{}_{he} \mathsf{r}^{\dagger}_{he})(E) =\sum_{\lambda, \mu} |(\mathsf{r}^{}_{he})_{\lambda,\mu}|^2(E)\quad,
\end{equation}
while the  normal $e \rightarrow e$ reflection is given by
\begin{equation}
R_{ee}(E) = {\rm Tr} (\mathsf{r}^{}_{ee} \mathsf{r}^{\dagger}_{ee})(E)  =\sum_{\lambda, \mu} |(\mathsf{r}^{}_{ee})_{\lambda,\mu}|^2(E)\quad.
\end{equation}
Moreover, one has  $R_{eh}(E)=R_{he}(-E)$  and  $R_{ee}(E)=R_{hh}(-E)$.
The total electron transmission coefficient from N to S is defined as
\begin{equation}
    T_e(E) = T_{ee}(E) + T_{he}(E)\quad,
\end{equation} 
where $T_{ee}(E)$ and $T_{he}(E)$  are defined as 
\begin{align}
   T_{ee}(E) &= {\rm Tr} (\mathsf{t}^{}_{ee} \mathsf{t}^{\dagger}_{ee})(E) =  \sum_{\lambda, \mu} |(\mathsf{t}^{}_{ee})_{\lambda,\mu}|^2(E)\quad \\
   T_{he}(E) &= {\rm Tr} (\mathsf{t}^{}_{he} \mathsf{t}^{\dagger}_{he})(E) = \sum_{\lambda, \mu} |(\mathsf{t}^{}_{he})_{\lambda,\mu}|^2(E) \quad,
\end{align}
and represent the total $e \rightarrow e$ and $e \rightarrow h$ transmission coefficients, respectively.

The nonlinear conductance at zero temperature is given by
\begin{equation}
\begin{split}\label{non-linear-conductance}
G(V)&=\frac{{\rm e}^2}{h}\left(\mathcal{N}_e({\rm e}V)-R_{ee}({\rm e}V)+R_{he}({\rm e}V)\right)\vartheta(V) \,+ \\
&+\frac{{\rm e}^2}{h}\left(\mathcal{N}_h(-{\rm e}V)-R_{hh}(-{\rm e}V)+R_{eh}(-{\rm e}V)\right)\vartheta(-V)  \quad,
\end{split}
\end{equation}
where $V$ is the voltage bias applied to the N-side as compared to the S-side, $\mathcal{N}_e(E),\mathcal{N}_h(E)\,$ are the number of electron and hole modes incoming  from  the lead N at energy $E$, respectively, and  $\vartheta(\cdot)$ denotes the Heaviside function.  \\

\subsection{Andreev reflection and conductance}
Let us now illustrate how the  AR coefficient and the conductance of the hybrid junction are affected by the relative   angle $\phi_{SO}$ of  the SO field direction. To this purpose, we set for definiteness the
same value of SO energies    $E_{SO,N}=E_{SO,S} = 0.5 \Delta_0$ in the two sides of the junction, corresponding to the same magnitude $\alpha=|\boldsymbol\alpha_N|=|\boldsymbol\alpha_S|$ [see Eq.(\ref{ESOj-def})].  Furthermore, we assume  a vanishing chemical potential in both side of the junction, i.e. $\mu_N = \mu_S = 0$, which corresponds to a Fermi energy lying in the middle of the magnetic gap, so that the critical Zeeman energy (\ref{EZc}) for the topological transition in the S-side acquires the value $E_Z^c = \Delta_0$. We shall compare the results for two sets of parameters, corresponding to the two cases where the proximitized  S-side of the NW is in the topological and in the trivial phase.
\begin{figure*}
    \centering
\includegraphics[width=1.0\linewidth]{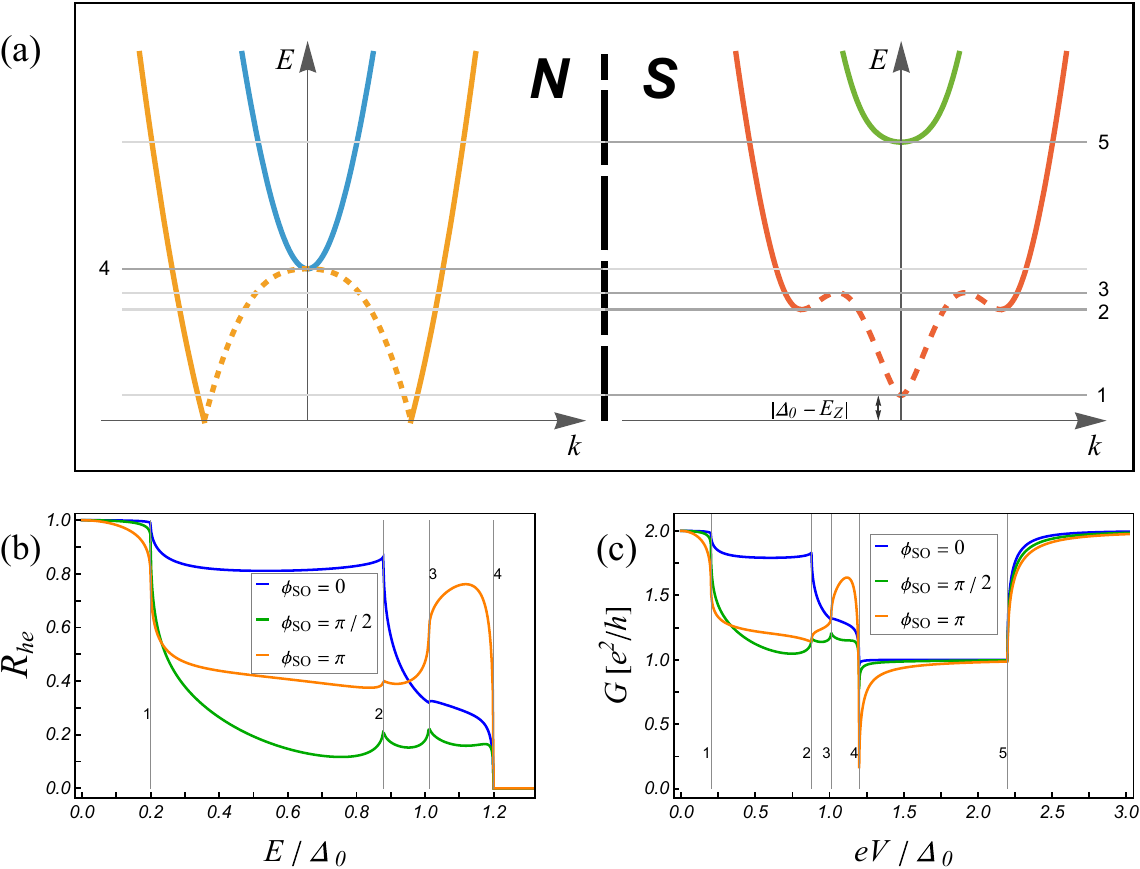} 
\caption{\label{Fig-2-Topological}The NW hybrid N-S junction for the parameters $E_{SO,N}=E_{SO,S} = 0.5 \Delta_0$,  $\mu_N = \mu_S = 0$ and $E_Z = 1.2 \Delta_0$. The S-side is in the topological phase and   the N-side is in the   Zeeman regime. (a) Bulk excitation spectra of the N  and S sides, where solid and dashed curves describe electron(like) and hole(like) excitations, respectively. The thin horizontal lines labeled from 1 to 5  identify the energy extrema  of the two bands, corresponding to $(E_1, E_2, E_3, E_4, E_5)=(0.20,0.88,1.01,1.20,2.20) \Delta_0$.
(b) The AR coefficient $R_{he}(E)$ as a function of the energy $E$, in units of the induced superconducting gap parameter $\Delta_0$, for three different values of the SO misalignment   angle,   $\phi_{SO} = 0, \pi/2, \pi$. The thin vertical lines are a guide to the  eye and correspond  to the energy values of the thin horizontal lines in panel (a).
At zero energy, AR is perfect, $R_{he}(E=0) = 1$, as a hallmark of the topological phase, and is also weakly dependent on $\phi_{SO}$  for $E \ll E_{1}$. In contrast, $R_{he}$  depends significantly on $\phi_{SO}$ in the energy interval $E_{1}<E< E_{4}$. For $E> E_{4}$, AR is suppressed because the N-side does not exhibit propagating  hole modes in such energy range. (c) The nonlinear conductance $G(V)$   in units of ${\rm e}^2/h$ as a function of the applied voltage bias $V$, in units of the induced superconducting gap parameter $\Delta_0$,  for   $\phi_{SO} = 0, \pi/2,\pi $.}
\end{figure*}

\subsubsection{S-side in the topological phase}
We start by analyzing the case where the S-side  is 
in the  topological phase ($E_Z > E_Z^c=\Delta_0$). Specifically, we take a Zeeman field $E_Z = 1.2 \Delta_0$, which also corresponds to the N-side in the Zeeman dominated regime. Figure~\ref{Fig-2-Topological}(a) shows the excitation spectra of the two sides of the junction, where thick solid and dashed curves denote electron(like) and hole(like) excitations, respectively, while the thin horizontal  lines are a guide to the eye to highlight the   energy extremal values $E_l$ of the spectra, labelled by $l=1 \ldots  {5}$. In particular, on the S-side of the junction [right-hand side of panel (a)], the  minimum of lower red band occurring at $k=0$ corresponds to the proximity superconducting gap $E_{1} = |\Delta_0-E_Z|=0.2 \Delta_0$, while the energy $  E_{5} = |\Delta_0+E_Z|=2.2 \Delta_0$  sets the opening of   propagating states of the upper   green band.

As mentioned earlier, the  spectra shown in Fig.\ref{Fig-2-Topological}(a) do not depend on the relative angle $\phi_{SO}$ of the SO field in the two sides.
The situation is different for the AR coefficient $R_{he}$, displayed in Fig.\ref{Fig-2-Topological}(b)  as a function of the energy $E$, in units of the induced gap parameter $\Delta_0$. The three curves refer to three different values $\phi_{SO}=0, \pi/2, \pi$ of the SO misalignment angle. The thin vertical lines are a guide to the  eye that correspond  to the energy values of the thin horizontal lines in panel (a). As one can see, when $E \rightarrow 0$, one has perfect  AR, $R_{he}=1$,  regardless of the specific value of $\phi_{SO}$. The robustness to perturbations of   $R_{he}(E=0)$ is a well known hallmark of the topological phase \cite{beenakker_prl_2011}, which persists even if, differently from an interface with vacuum,  at the N-S interface the Majorana bound state no longer exists as an isolated state, as it hybridizes with the propagating states of the continuum spectrum of the N-side~\cite{Wimmer_njp_2011}.  

For finite but small energy, $E \ll \Delta_0$, the dependence of $R_{he}$ on $\phi_{SO}$ remains very weak. Only for higher energy values, namely for $E_{1}<E<E_{4}$, i.e. when propagating states are present  also in the proximitized side, AR is significantly affected by the misalignment angle. Then, for $E>E_{4}=1.2 \Delta_0$, AR is suppressed because there are no hole states in the excitation spectrum on the N-side.

Fig.\ref{Fig-2-Topological}(c) shows the behavior of the nonlinear conductance (\ref{non-linear-conductance}) of the hybrid NW junction as a function of the voltage $V$ applied to the N-side, again for the three different values of the misalignment angle $\phi_{SO}$. As one can see, the dependence of $G$ on $\phi_{SO}$, especially in the voltage range $E_{1}<{\rm e}V<E_{2}$ straightforwardly reflects  the dependence of the  AR process on  $\phi_{SO}$ shown in panel (b). Note, in particular, the perfectly quantized  value   $G(V=0)=2 {\rm e^2}/h$ at zero-bias~\cite{beenakker_prl_2011,Wimmer_njp_2011}.
Furthermore, at finite voltage, $G(V)$ also exhibits cusps at the energy values corresponding to the local minima highlighted by thin lines in panel (a). The cusp at ${\rm e}V=E_{4}$ is particularly pronounced in the case of complete misalignment  angle, $\phi_{SO}=\pi$, where the SO field has opposite directions in the N and S-sides. This corresponds to the transition where no hole states are present on the N-side, and AR is suppressed for ${\rm e}V>E_{4}$. Finally, for ${\rm e}V>E_{5}$, the conductance exhibits a jump due to the onset of the upper band propagating channels [green curve of panel (b)]. In this regime of propagating states, the dependence on $\phi_{SO}$ becomes very weak.

\subsubsection{S-side in the trivial phase}
\begin{figure*}
    \centering
    \includegraphics[width=1.0\linewidth]{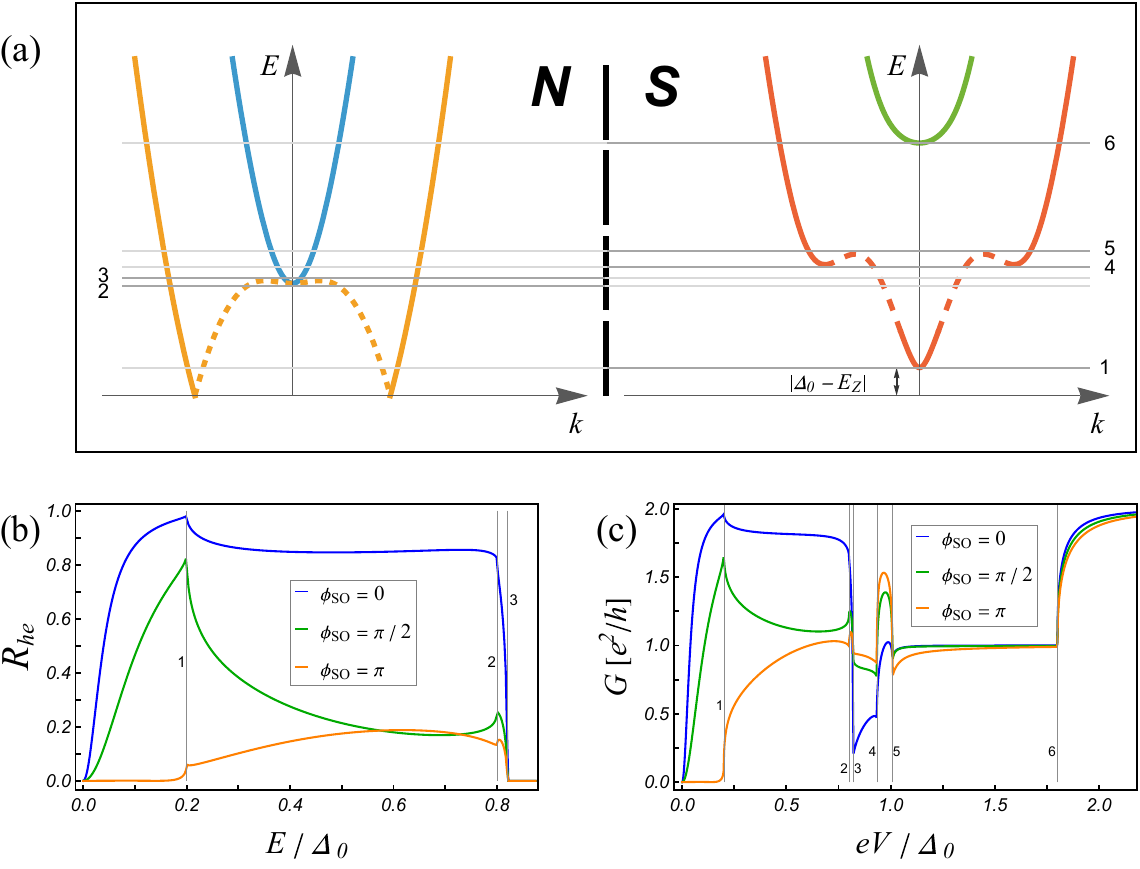} 
   \caption{\label{Fig-3-Trivial}The NW hybrid N-S junction for the parameters $E_{SO,N}=E_{SO,S} = 0.5 \Delta_0$,  $\mu_N = \mu_S = 0$ and $E_Z = 0.8 \Delta_0$. The S-side is in the trivial phase, while   the N-side is in the   Rashba dominated regime.   
   (a) Bulk excitation spectra of the N and S sides (notation like in Fig.\ref{Fig-2-Topological}). Here, the horizontal lines correspond to the energy band extrema  $(E_1, E_2, E_3, E_4, E_5,E_6)=(0.20,0.80,0.82,0.93,1.01,1.80) \Delta_0$.  
(b) The  AR coefficient $R_{he}(E)$  for three different values of the SO misalignment angle,   $\phi_{SO} =0, \pi/2,\pi$. At zero energy, $R_{he}(E=0) = 0$ since the S-side is in the trivial regime. At finite energy, $R_{he}(E)$ exhibits a strong dependence on $\phi_{SO}$ for $0<E< E_{1}$. For $E=0.1 \Delta_0$,  one can tune its value from $R_{he} \simeq 0.87$   (for $\phi_{SO}=0$) down to  $R_{he} \simeq 4 \cdot 10^{-4}$ (for  $\phi_{SO}= \pi$).   (c) Nonlinear conductance $G(V)$  in units of $e^2/h$ as a function of the applied voltage bias $V$, for $\phi_{SO}  =0,\pi/2,\pi$. Similarly to $R_{he}(E)$, also $G(V)$ exhibits a strong $\phi_{SO}$-dependence. }
\end{figure*}

Let us now come to the case where the S-side  is in the topologically trivial regime ($E_Z < E_Z^c=\Delta_0$). Specifically,    we analyze the case of a Zeeman energy $E_Z = 0.8\, \Delta_0$. The bulk excitation spectra of the N  and S sides are shown Fig.\ref{Fig-3-Trivial}(a), where    the horizontal thin lines {1} to {6} again identify the energy extrema of the bands, at which a change in the number of propagating modes occurs.
As compared to the case of the topological phase, the weaker Zeeman field ($E_Z < E_Z^c$) causes the N-side to be now  in the Rashba dominated regime. This can be appreciated by the appearance of  a mexican hat shape in its hole spectrum [yellow dashed curve on the left-hand side of Fig.\ref{Fig-3-Trivial}(a)], as highlighted by the two horizontal thin lines {2} and {3}   corresponding to the energies $E_{2}=E_Z- \mu_N = 0.8 \Delta_0$ of the local minimum at $k=0$, and $E_{3}= \mu_N + E_{SO,N}+ E_Z^2/4E_{SO,N} =0.82 \Delta_0$ of the two local maxima of the hole spectrum at $k= \pm k_{SO}\sqrt{1-E_Z^2/4 E_{SO}^2}$, respectively, with $k_{SO} = \sqrt{2m^*E_{SO,N}}/\hbar$.
As far as the spectrum of the S-side    of the junction is concerned [right-hand side of Fig.\ref{Fig-3-Trivial}(a)], only minor quantitative changes  from Fig.\ref{Fig-2-Topological}(a) arise, due   to the different value of the Zeeman energy $E_Z$. Thus,    $E_{1} = | \Delta_0 -E_Z| = 0.20 \Delta_0$, $E_{4}\simeq 0.93 \Delta_0$, $E_{5} \simeq 1.01\Delta_0$, while $E_{6} =   | \Delta_0 + E_Z| = 1.80 \Delta_0$.

More relevant  differences from the topological phase   appear when considering the   behavior of the AR coefficient $R_{he}(E)$ as a function of energy, shown in Fig.\ref{Fig-3-Trivial}(b).  While for $E \rightarrow 0$ the reflection coefficient is completely suppressed ($R_{he}(0)\rightarrow 0$), in agreement with the result of Ref.\cite{beenakker_prl_2011} for the trivial phase of S,
a relatively small finite energy $E \lesssim E_{1}$ is sufficient to give rise to a strong dependence on the SO misalignment angle $\phi_{SO}$. Indeed the three displayed colored curves indicate that, already for an energy as small as $E=0.1\Delta_0$, the AR coefficient $R_{he}$  can vary from a high value $R_{he} \simeq 0.87$ for aligned SO directions ($\phi_{SO}=0$), to an intermediate value $R_{he} \simeq 0.47$  (for $\phi_{SO}=\pi/2$), down to a complete AR suppression  $R_{he} \simeq 4 \cdot 10^{-4}$ for a full misalignment ($\phi_{SO}= \pi$).
Again, the thin vertical lines of Fig.\ref{Fig-3-Trivial}(b) are a guide to the eye and correspond to the energy values highlighted by the horizontal lines of the spectra in panel (a). The tunability of the AR is still present in the range $E_{1}< E < E_{3} = 0.82 \Delta_0$, whereas 
for   $E > E_{3}$,   AR is completely suppressed for any angle $\phi_{SO}$ because no hole excitations are present on the N-side, as can be inspected from Fig.\ref{Fig-3-Trivial}(a).

The tunability of the AR also reflects onto the nonlinear conductance $G(V)$, shown in  
Fig.\ref{Fig-3-Trivial}(c). Notice, in particular, how the misalignment angle $\phi_{SO}$ enables one to tune the conductance $G$ at low-voltage ${\rm e}V<E_{1}$, in striking contrast to what happens in the topological phase case, where it is pinned near $G\simeq {\rm e^2}/h$.   
We also observe that, although     AR is completely suppressed for   $E > E_{3}$ [Fig.\ref{Fig-3-Trivial}(b)], the conductance still depends on $\phi_{SO}$ in the range  $E_{3}<E < E_{5}$, due to the normal reflection contribution $R_{ee}$ to the conductance [see Eq.\eqref{non-linear-conductance}]. Above the superconducting gap parameter, ${\rm e}V>\Delta_0$, the dependence on $\phi_{SO}$ becomes negligible.



\section{Discussion}
\label{sec-4}
In this section we would like to discuss a few aspects related to the results presented above. In particular, we shall address the origin of the tunability of the AR,  and analyze the dependence of the nonlinear conductance on the Zeeman field~$E_Z$. Then, we shall discuss some possible realistic setup implementations where the predicted effects can be observed.

\subsection{Origin of the tunability of the conductance}
In  section \ref{sec-3} we have shown that, although the bulk spectra of the N-side and S-side  are insensitive to the orientation of the SO field, the AR coefficient and   the nonlinear conductance of the hybrid N-S junction   strongly depend  on  the SO misalignment angle~$\phi_{SO}$. Here, we would like to comment about the origin of such an effect. Although a thorough explanation requires the 
numerical analysis of all scattering states  at every energy~$E$,  one can still identify some general features that help understand the origin of this phenomenon.

We start by observing that the boundary conditions (\ref{bc-fun})-(\ref{bc-der}) at the N-S interface depend explicitly only on the two SO fields $\boldsymbol{\alpha}_N$ and $\boldsymbol{\alpha}_S$, and not on  the magnetic field $\mathbf{b}$. Thus, one might at first naively expect that, even for $\mathbf{b}=0$, $R_{he}(E)$ and $G(V)$ could be tuned with~$\phi_{SO}$. However, this is not the case: without an applied magnetic field  the AR coefficient and   the nonlinear conductance are completely {\it insensitive} to the misalignment angle $\phi_{SO}$ between the N and S sides, just like the bulk spectra.
Indeed, if $\mathbf{b}=0$,  on each  junction side $L=N,S$, the bulk Hamiltonian commutes with the spin-operator $\boldsymbol{\sigma}\cdot \boldsymbol{\alpha}$ related to the SO field, 
\begin{equation}\label{symm-b=0}
 \left[  \left.  H_{{\rm BdG},L}\right|_{\mathbf{b}=0} \,,\, 
\tau_0 \otimes (\boldsymbol{\sigma}\cdot \boldsymbol{\alpha}_L)\right]=0 \hspace{0.8cm} L=N,S\quad, 
\end{equation}
implying that all   states can be classified as parallel or antiparallel to the direction $\boldsymbol{\alpha}_L$, i.e. according to the eigenvalue $\lambda_L=\pm 1$ of $\boldsymbol{\sigma}\cdot \hat{\boldsymbol{\alpha}}_L$, with $\hat{\boldsymbol{\alpha}}_L$ being the unit vector related to $\boldsymbol{\alpha}_L$.
Thus, in an AR process in the absence of magnetic field, an electron impinging from the N-side  with  spin    (say) parallel to $\hat{\boldsymbol{\alpha}}_N$ is back-reflected as a hole with an antiparallel spin  $-\hat{\boldsymbol{\alpha}}_N$, giving rise to a spin-$\hat{\boldsymbol{\alpha}}_N$ excitation at energy $E>0$. Moreover, on the N-side
there are always {\it two} incoming  electron states, with opposite spin orientations, parallel and antiparallel  to $\hat{\boldsymbol{\alpha}}_N$ and, correspondingly,  {\it two} outgoing hole  states with spin orientations   antiparallel and parallel to~$\hat{\boldsymbol{\alpha}}_N$. A similar correspondence occurs for electron-like and hole-like states on the S-side, where excitations can be classified as parallel and antiparallel to $\hat{\boldsymbol{\alpha}}_S$.
Thus, although N and S sides are characterized by  different spin orientations $\boldsymbol{\alpha}_N$ and $\boldsymbol{\alpha}_S$, the excitations  always form a complete basis for the spin Hilbert space. As a consequence, in the absence of a magnetic field, the total AR process is not physically affected by the misalignment angle $\phi_{SO}$, which   simply corresponds to a mere basis rotation in the spin Hilbert space.

However,  when a magnetic field $\mathbf{b}$ is applied  perpendicular to the SO directions $\boldsymbol{\alpha}_N$ and $\boldsymbol{\alpha}_S$,  the symmetries in Eq.(\ref{symm-b=0}) get broken. In fact, the magnetic field has a twofold effect. First, it provides a spin texture to the  NW states, i.e. the spin direction of each $k$-state acquires a dependence on its wavevector $k$~\cite{streda_2003,houzet-meyer_2016,rossi-dolcini_prb_2018}. For instance, on the N-side, the spin of an electron/hole  state experiences an effective magnetic field $\mathbf{b}_{eff}=\mathbf{b}\pm k\boldsymbol{\alpha}_N$,  
implying that the spin directions of the impinging electron and the Andreev reflected hole are no longer antiparallel when $\mathbf{b} \neq 0$. Notice that such spin orientation mismatch increases with the excitation energy $E$, since the  difference between the values $k_{N,e}$ and $k_{N,h}$ of the electron and hole wavevectors does. In order for the AR to occur, the spin orientation mismatch  between electron and hole in the N-side must be supplied at the interface by the spin orientation of the wavefunction  on  the S-side. The difference in its SO field direction~$\boldsymbol{\alpha}_S$ with respect to ~$\boldsymbol{\alpha}_N$ affects the AR process.

The second effect of the magnetic field is to lead to an energy separation between the two bands characterizing each side of the junction, as can clearly be seen from the spectra depicted in Figs.\ref{Fig-2-Topological}(a) and \ref{Fig-3-Trivial}(a), where   yellow and blue bands (on the N-side) and red and green bands (in the S-side) are energetically separated from each other.  In an energy range where only one band  is involved, 
the spin of its states no longer form a complete basis of the spin Hilbert space. For instance, for an excitation energy $0<E<E_{2}$ in Fig.\ref{Fig-3-Trivial}(a), there is only {\it one} right-moving electron  state impinging from the N-side to the interface, and only {\it one} left-moving hole state, both originating from the lower band ($b=-$) depicted in yellow. Similarly, on the S-side, the evanescent states only originate from the  lower red band ($b=-$) not only for energies  below the gap $E_{1}=|\Delta_0-E_Z|=0.2 \Delta_0$, but also in the range $0< E\lesssim E_{3}$.  
\begin{notainterna}
\leo{Strettamente parlando, $E_3 = 0.82\Delta_0$ è un'energia di soglia della regione N (non S). In ogni caso, confermo che fino a circa $\approx 0.822\Delta_0$ gli stati evanescenti della regione S provengono tutti dalla banda rossa}  
\end{notainterna}
As a consequence, when such a lack of spin degrees of freedom  caused by the magnetic field is present on both sides of the interface, a difference by an angle $\phi_{SO}$ in the SO directions $\boldsymbol{\alpha}_N$ and $\boldsymbol{\alpha}_S$  does not simply amount to a spin basis rotation. Rather,  for the propagating states
the interface acts as a spin polarization filter, and the AR coefficient does depend on the misalignment angle~$\phi_{SO}$.

For these reasons, the presence of a nonvanishing magnetic field is crucial for the tunability with $\phi_{SO}$. However, a too strong magnetic field $\mathbf{b}$ tends align the electron and hole states either parallel or antiparallel to $\mathbf{b}$, irrespective to the directions $\boldsymbol{\alpha}_N$ and $ \boldsymbol{\alpha}_S$. 
As a matter of fact, the critical Zeeman energy $E_Z^c$  Eq.(\ref{EZc}) identifying the transition from the topological to the trivial phase of the S-side also  roughly separates the crossover from a weak to a strong tunability of AR and conductance at low energies, as can be seen 
by comparing Figs.\ref{Fig-2-Topological}  and \ref{Fig-3-Trivial}.

In particular, the strong suppression observed in $R_{he}$ and $G(V)$ in Fig.\ref{Fig-3-Trivial}(b) and (c) at (say) $E={\rm e}V=0.1 \Delta_0$ when  $\phi_{SO}$ is increased from 0 to $\pi$ can be qualitatively understood by inspecting the second boundary conditions Eq.(\ref{bc-der}). Indeed, for $\phi_{SO}=0$, i.e. for $\boldsymbol{\alpha}_N=\boldsymbol{\alpha}_S$,   Eq.(\ref{bc-der}) implies that the wavefunction spatial derivative  is continuous at the interface. This favors processes with equal wavevectors, such as an  electron impinging from N with wavevector $k_{N,e}$ being back-reflected as a hole with similar wavevector $k_{N,h}  \simeq  k_{N,e}$. By contrast,   a misalignment angle $\phi_{SO}$ between   $\boldsymbol{\alpha}_N$ and $\boldsymbol{\alpha}_S$ causes a discontinuity in the derivative, opening up the possibility of processes with a momentum mismatch, such as an incident electron with wavevector $k_{N,e}$  experiencing also normal reflection with opposite wavevector $-k_{N,e}$. When $\phi_{SO}=\pi$ (antiparallel SO configuration between N and S), the AR is completely suppressed in favor of normal reflection.

\subsection{Conductance dependence on $E_Z$ and $\phi_{SO}$}
\begin{figure}
    \centering
    \includegraphics[width=1.0\linewidth]{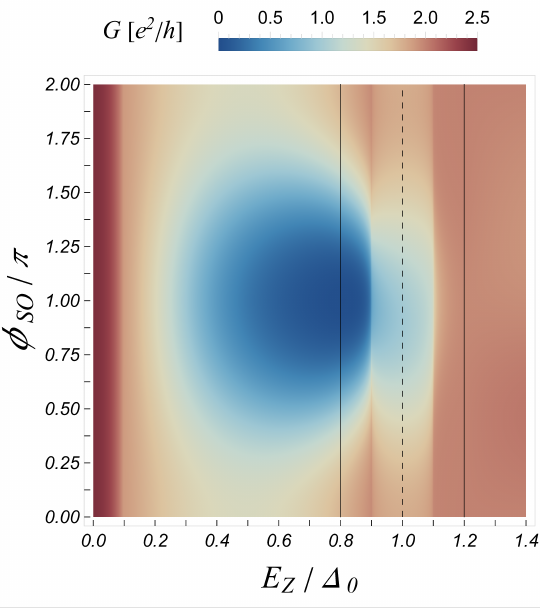}
    \caption{\label{Fig-4-densityplot}  Density plot of the nonlinear  conductance $G(V)$, in units of ${\rm e}^2/h$, at fixed voltage ${\rm e}V = 0.1 \Delta_0$, shown as a function of the Zeeman energy $E_Z$ (in terms of the superconducting gap parameter $\Delta_0$) and of the $\phi_{SO}$ misalignment angle. The other energy parameters are set to $E_{SO,N}=E_{SO,S} = 0.5 \Delta_0$ and $\mu_N=\mu_S = 0$. The vertical dashed black line marks the critical Zeeman energy value $E_Z^c = \Delta_0$ separating the trivial from the topological phase for the S-side, while the two vertical solid black lines at $E_Z = (0.8,1.2)\,\Delta_0$ identify the   two cuts  analyzed in Figs.~\ref{Fig-3-Trivial} and \ref{Fig-2-Topological}, respectively. The figure showcases the high tunability of the conductance in the topological trivial regime, particularly for $0.4 \le E_Z/\Delta_0 \le 0.9$, whereas~$G$ is  poorly tunable  in the topological regime. }
\end{figure}

The results about the  nonlinear conductance $G(V)$ presented above have revealed that, at finite but small voltage ${\rm e}V \ll \Delta_0$,  the Zeeman field plays a quite relevant role in determining the  tunability of $G(V)$ as a function of the misalignment angle $\phi_{SO}$.  This  is better highlighted in the density plot shown in Fig.\ref{Fig-4-densityplot}, where   the low-voltage conductance, i.e. for ${\rm e}V=0.1 \Delta_0$, is shown as a function of the applied Zeeman field $E_Z$ and of the misalignment SO angle $\phi_{SO}$. 

For a Zeeman field smaller than the applied voltage, $E_Z < {\rm e}V=0.1 \Delta_0$, the conductance     is independent of $\phi_{SO}$, as can be inspected by the dark red vertical stripe on the left of Fig.\ref{Fig-4-densityplot}, where $G$ exceeds the single-channel value $G=2{\rm e^2}/{h}$ of  a N-S junction. Indeed, at such a weak field, the energy $E={\rm e} V$ lies above the magnetic gap  and  two propagating channels are available on  the N-side   (two incoming electron states and two outgoing  hole states).  All the spin degrees of freedom are available on both sides of the junction and, as observed above, the misalignment angle $\phi_{SO}$ simply amounts to a basis rotation, with no physical effects. 

However, for $E_Z> {\rm e}V=0.1 \Delta_0$,   the energy falls inside the magnetic gap, and only a single transport channel remains open, leading to a   cusp-like decrease of the conductance. In turn, this leads to a decrease in the available spin degrees of freedom, as observed above, and the dependence on the misalignment angle $\phi_{SO}$ starts to emerge. It becomes   more pronounced for $0.5 \lesssim  E_Z/\Delta_0  \lesssim 0.9$, where the states on both N and S sides  contain only  half of the complete spin basis. Thus, varying the $\phi_{SO}$ angle affects the spin-matching between the wavefunctions of the two sides. The vertical line at $E_Z=0.8 \Delta_0$ represents the cut related to Fig.\ref{Fig-3-Trivial}(c), where the S-side is in the trivial phase: The conductance   $G$ can be tuned from values as low as $10^{-3}$
up to $\approx 1.74$ (in units of ${\rm e^2}/h$), with varying the SO angle $\phi_{SO}$.

Furthermore, a change 
 in the behaviour of the conductance is also observed in the energy Zeeman range $E_Z \in [0.9,1.1]\Delta_0$. At the two extremal values of such range, the applied voltage ${\rm e}V=0.1\Delta_0$ exactly matches the bulk superconducting gap $|E_Z-\Delta_0|$. 
Thus, within such a range of Zeeman field values, the transmission process from the N- to the S-side of the junction opens up. This causes a decrease in the Normal Reflection [$R_{ee}(eV)$ in Eq.\eqref{non-linear-conductance}] and consequently an increase in the conductance, which becomes less tunable.
Notice that the middle value $E_Z=\Delta_0$ of such a range corresponds to the   transition of the S-side from the trivial to topological phase.

Finally, when $E_Z > 1.1\Delta_0$,  transport  occurs entirely below the superconducting gap. In particular, the vertical line at $E_Z/\Delta_0=1.2$ represents the cut related to Fig.\ref{Fig-2-Topological}(c). As observed above,  large values of Zeeman field tend to polarize the  states spin, and the tunability of the conductance $G$ with respect to the SO angle $\phi_{SO}$ is almost entirely suppressed. \\

\subsection{Possible implementations of the setup}
Let us now discuss some possible realistic setups where the predicted results about conductance could be observed. NWs  based on InSb or InAs   characterized by a ballistic transport and a strong SO coupling  have been fabricated in a number of experiments~\cite{heiblum_2012,kouwenhoven_science_2012,furdyna_natphys_2012,xu_2012,nilsson_2012,marcus_2016,kouwenhoven_nanolett_2017,kouwenhoven_natnano_2018}. 
Proximization   has been realized by depositing a superconducting film of Al \cite{heiblum_2012,nilsson_2012,marcus_2016}, Nb~\cite{furdyna_natphys_2012,xu_2012} or NbTiN 
\cite{kouwenhoven_science_2012,kouwenhoven_nanolett_2017,kouwenhoven_natnano_2018}  on a NW segment. 
Focusing e.g. on InSb,  the  effective electron mass is $m^* = 0.015\,m_e$, and the typical value of SO energy is $E_{SO} \sim 0.05 \, {\rm meV}$~\cite{kouwenhoven_science_2012}. With NbTiN films the induced gap parameter is on the order of $\Delta_0 \simeq 0.09 - 0.25 {\rm meV}$~\cite{kouwenhoven_science_2012,kouwenhoven_natnano_2018}. The g-factor is $g^* \approx 50$~\cite{fan_2015}, and an applied magnetic field of $B \sim 50 - 100 \, {\rm mT}$, leads to a Zeeman energy $E_Z = {g} \mu_B B/2$ on the order of $E_\mathrm{Z} \sim 0.07 - 0.14 \, {\rm meV}$.  The energy ratios used in our plots are compatible with these experimental ranges.

The magnitude of the SO coupling can be controlled via electrolyte gating~\cite{gao_2012}, local top gates~\cite{kouwenhoven-wimmer_2015}  or bottom gates patterned beneath the NW~\cite{kouwenhoven-natcomm_2017}  or also with  gate-all-around geometries, where the NW is fully enclosed by a metallic gate separated by a dielectric 
layer~\cite{sasaki_2013,sasaki_2017,sasaki_2021,guo_2021}.  
 Moreover, the large variety of  realized gate geometries  enable one to  control also the direction of the SO field on the N-side of the junction.  Side gates, for instance, introduce lateral electric fields that break left--right symmetry and enable rotation of the effective spin--orbit field~\cite{nygard_2016,strambini_2018,burke_2019}. Top gates can be combined with bottom or side gates to further enhance tunability. Wrap-around gates, also referred to as $\Omega$-gates, partially surround the nanowire and provide strong capacitive coupling~\cite{moon_2006,wernersson_2008, wernersson_2010,deshmukh_2011,micolich_2015}. Furthermore,  lateral deposition of the superconducting film proximitizing a NW portion   also modifies the
  SO field direction  on   the S-side~\cite{kouwenhoven_prl_2019,kouwenhoven_prappl_2024}.
All these architectures permit precise control of the Rashba coupling, and represent a viable route to observe the predicted tunability effect.

\section{Conclusions}
\label{sec-5}
In this paper, we have investigated how the AR coefficient $R_{he}(E)$ and the nonlinear conductance $G(V)$ in a hybrid N-S junction based on a NW depend on the different orientations $\boldsymbol{\alpha}_N$ and $\boldsymbol{\alpha}_S$  of the SO field on the two sides of the junction. While the energy spectra in the bulk of the two sides are independent of the SO field direction and only depend on the SO magnitude $|\boldsymbol{\alpha}|$, we have shown that in  the hybrid N-S junction    SO directions do  matter. 
Indeed, when a magnetic field $\mathbf{b}$ is applied along the axis of the NW, i.e. perpendicularly to the SO field, the misalignment angle $\phi_{SO}$ between the two SO field directions $\boldsymbol{\alpha}_N$ and $\boldsymbol{\alpha}_S$ strongly impacts on the Andreev energy-dependence of the reflection coefficient $R_{he}(E)$ and the nonlinear conductance $G(V)$ behavior as a function of the applied voltage bias $V$. 

At zero-energy, we have recovered the expected behavior of the AR and linear conductance, which are determined by the topological phase of the S-side of the junction: In the trivial phase one has  $R_{he}=0=G(V=0)$, whereas in the topological phase one finds $R_{he}=1$ and $G(V=0)=2{\rm e^2}/h$, regardless of the value of the SO misalignment $\phi_{SO}$. Such a zero-energy robustness to the SO direction generalizes the results   found in N-S junctions in the presence of disorder~\cite{beenakker_prl_2011,Wimmer_njp_2011}.

However, at finite energy, the SO direction misalignment becomes important. In particular, at finite but low energies $0< E,{\rm e}V \lesssim 0.2 \Delta_0$, $R_{he}(E)$ and $G(V)$ remain  close to the zero energy value and are poorly tunable with $\phi_{SO}$, if the S-side is in the topological phase [see Fig.\ref{Fig-2-Topological}(b) and (c)]. In striking contrast, when the S-side is in the trivial phase, a wide tunability of  $R_{he}(E)$ and $G(V)$ as a function of the SO misalignment is observed at finite energy. As shown in Fig.\ref{Fig-3-Trivial}(b), the AR coefficient $R_{he}$  can be tuned from a high value $R_{he} \simeq 0.87$ for aligned SO directions ($\phi_{SO}=0$), to an intermediate value $R_{he} \simeq 0.47$  (for $\phi_{SO}=\pi/2$), down to a complete AR suppression  $R_{he} \simeq 4 \cdot 10^{-4}$ for a full misalignment ($\phi_{SO}= \pi$). Similarly, $G({\rm e}V=0.1 \Delta_0)$ spans from $\approx 10^{-3}$ up to $\approx 1.74$ (in units of ${\rm e^2}/h$), with varying the SO angle $\phi_{SO}$. 

We have pointed out the crucial role played by the magnetic field on determining such a tunability, as highlighted in   Fig.\ref{Fig-4-densityplot}. At too weak magnetic field $E_Z < {\rm e}V$ the conductance is independent of $\phi_{SO}$ because this is the regime where AR is carried by two independent spin channels and   a misalignment of SO directions simply amounts to a spin basis rotation. Similarly, at too large magnetic field, $E_Z \gtrsim   \Delta_0$, the tunability with $\phi_{SO}$ is poor, for the magnetic field spin polarizes the two sides. In contrast, at intermediate Zeeman field values, particularly for $0.4 \le E_Z/\Delta_0 \le 0.9$, i.e. when the S-side is in the trivial phase, the conductance $G$ becomes strongly dependent on the misalignment angle. Finally, we have discussed the implementation in realistic NW-based setup, where SO direction can by now be controlled through various geometrical gating configurations and film deposition designs.

In conclusion,   while so far 
most studies  have focussed on controlling   electron transport  by enhancing  the range of the SO coupling magnitude in NWs, our results demonstrate that, even at fixed SO magnitude,  the SO direction is a quite efficient   knob to electrically tune the AR and the conductance of a   NW-based N-S hybrid quantum device. These findings, which could also have potential developments in NW-based Josephson junctions and   Andreev spin qubits, suggest that NWs are promising and versatile quantum systems even beyond the search for Majorana quasi-particles.  
 

\begin{acknowledgments}
The authors greatly acknowledge interesting discussions with F.~Taddei and A.~Braggio. F.D. also acknowledges  financial support from the  
MUR-PRIN 2022—Grant No. 2022B9P8LN-(PE3) Project NEThEQS “Non-equilibrium coherent thermal effects in quantum systems” in PNRR Mission 4-Component 2-Investment 1.1 “Fondo per il Programma Nazionale di Ricerca e Progetti di Rilevante Interesse Nazionale (PRIN)” funded by the Next Generation EU initiative, as well as from the TOPMASQ ("Topological material platform for the implementation of Andreev spin qubit") project,   a Cascade call project (CUP E13C24001560001) financed by the “National Quantum Science \& Technology Institute”, PE00000023 (Next Generation EU).

\end{acknowledgments}

\appendix

\section{Eigenstates in the bulk of the N and S sides of the junction}
\label{AppA}
Since in each  side the parameters are spatially uniform, one can build up the solution in the hybrid junction by matching the bulk solution in the bulk of the N-side  and in the bulk of the S-side. Here below, we specify some details about the bulk eigenstates of the two sides.  
\subsection{Solutions in the N-side}
In the N-side of the junction the  superconducting pairing vanishes, $\Delta_0 = 0$.
By re-expressing the Nambu spinor field operator    in Eq.\eqref{Nambu-spinor-real-space} as 
$\Psi(x)  = 
\Omega^{-1/2}\sum_k e^{i k x} \Psi_{k}$, 
where $\Psi_k=( 
{c}^{}_{k\uparrow},
{c}^{}_{k\downarrow},
{c}^\dagger_{-k\downarrow},
-{c}^\dagger_{-k\uparrow}
 )^T$ are the Nambu mode operator and $\Omega\rightarrow \infty$ is the entire length of the system, the Hamiltonian (\ref{H-NW-inhomo-1}) acquires a diagonal form in $k$-space,
$
\mathcal{H}=(1/2) \sum_k \Psi^\dagger_k H_{BdG}(k) \Psi_k^{} 
$,
where 
\begin{eqnarray} 
\lefteqn{H_{BdG}(k) =} \label{HBdG(k)-N} & & \\
& & 
\begin{pmatrix}
     \xi_N(k) \sigma_0  +\mathbf{m}_N(k)\cdot \boldsymbol{\sigma}    &   0
     \\ 0
     &  -\xi_N(k) \sigma_0 + \mathbf{m}_N(-k)\cdot \boldsymbol{\sigma} 
\end{pmatrix} \nonumber \quad.
\end{eqnarray} 
Here, $\xi_N(k) =\hbar^2 k^2/2 m^* -\mu_N$ and
\begin{eqnarray}\label{m(k)-def}
\mathbf{m}_N(k)&=&- (  \boldsymbol{\alpha}_N k + \mathbf{b}) \\&=&\left(-b_x\, , -\alpha_N k\, \sin\phi_{SO}\, ,\, -\alpha_N k\cos\phi_{SO}\, \right)\nonumber
\end{eqnarray}
is a $k$-dependent vector determined by the SO field (\ref{alphaN-N-vec-def}) and the magnetic field $\mathbf{b}$.
Because the BdG Hamiltonian (\ref{HBdG(k)-N}) in the N-side naturally decouples into its two $2\times 2$ electron and hole  blocks, its diagonalization is straightforward and enables one to directly identify, for each of the two excitation bands $E_{N,\pm}(k)$ in Eq.(\ref{spectrum-N-side}), the electron(e) and hole(h)  branches.\\

\noindent The {\it electron branch} corresponds to the case $\xi_N(k) \pm \sqrt{b_x^2+\alpha_N^2 k^2} >0$. It has a spectrum 
\begin{equation}
E_{N,e,\pm}(k) =\vartheta\left(\xi_N(k) \pm \sqrt{b_x^2+\alpha_N^2 k^2}  \right) E_{N,\pm}(k)  \quad,\label{spectrum-N-side-e-branch}
\end{equation}
and its corresponding eigenvectors  are
\begin{equation}\label{2-eigenvectors-N-side-e}
\begin{array}{l}
\mathsf{w}_{N,e,+}(k)  = 
    \Big( \cos\frac{\theta_k}{2}, \, \sin\frac{\theta_k}{2}\, e^{i\varphi_k}, \, 0, \, 0 \Big)^T  \\
\mathsf{w}_{N,e,-}(k)  = 
    \Big( -\sin\frac{\theta_k}{2}\, e^{-i\varphi_k}, \, \cos\frac{\theta_k}{2}, \, 0, \, 0 \Big)^T  
\end{array}\quad.
\end{equation}

\noindent The {\it hole branch} corresponds to $\xi_N(k) \pm \sqrt{b_x^2+\alpha_N^2 k^2} <0$, has a  spectrum  
\begin{equation}
E_{N,h,\pm}(k) =\vartheta\left(-\xi_N(k) \mp \sqrt{b_x^2+\alpha_N^2 k^2}  \right) E_{N,\pm}(k)  \quad,\label{spectrum-N-side-h-branch}
\end{equation}
and eigenvectors  
\begin{equation}\label{2-eigenvectors-N-side-h}
\begin{array}{l}\mathsf{w}_{N,h,+}(k)  = 
    \Big( 0, \, 0, \, -\sin\frac{\theta_{-k}}{2}\, e^{-i\varphi_{-k}}, \, \cos\frac{\theta_{-k}}{2} \Big)^T   \\
\mathsf{w}_{N,h,-}(k)  = 
    \Big( 0, \, 0, \, \cos\frac{\theta_{-k}}{2}, \, \sin\frac{\theta_{-k}}{2}\, e^{i\varphi_{-k}} \Big)^T 
    \end{array} \quad.
\end{equation}
In Eqs.(\ref{2-eigenvectors-N-side-e}) and (\ref{2-eigenvectors-N-side-h}),  $\varphi_k \in[0, 2\pi]$ and $\theta_k\in[0,  \pi]$ are the azimuthal and polar angles characterizing the  
 unit vector of $\hat{\mathbf{m}}(k)$ in Eq.(\ref{m(k)-def})
\begin{equation}
\label{hat{n}(k)-def}
\hat{\mathbf{n}}_N(k)=\frac{\mathbf{m}_N(k)}{|\mathbf{m}_N(k)|}=  \left(\sin\theta_k \cos\varphi_k , \sin\theta_k \sin\varphi_k , \cos\theta_k   \right)\quad,
\end{equation}
and fulfill the relations 
\begin{eqnarray}
\cos\theta_k &=& \displaystyle -\frac{\alpha_N k \cos \phi_{SO}}{\sqrt{b_x^2+(\alpha_N k)^2}} \label{Spherical-coord-costheta} \\
&=& \displaystyle   - \frac{2  \mbox{\small sgn}(k)   \cos \phi_{SO}\sqrt{\varepsilon_0(k)E_{SO,N}}}{\sqrt{E_Z^2+4 \varepsilon_0(k)E_{SO,N}}} \nonumber \\& & \nonumber\\
\sin\theta_k &=& \displaystyle \sqrt{\frac{b_x^2+(\alpha_N\, k\sin{\phi_{SO}} )^2}{b_x^2+(\alpha_N k)^2}} \label{Spherical-coord-sintheta} \\
    &=&  \displaystyle \sqrt{\frac{E_Z^2+4 \varepsilon_0(k) E_{SO,N}\sin^2{\phi_{SO}} }{E_Z^2+4 \varepsilon_0(k)E_{SO,N}}}
    \nonumber \\ & & \nonumber\\
\varphi_k &=&  \displaystyle\arctan\left( \frac{\alpha_N\, k\sin{\phi_{SO}} }{ b_x}\right) \,+\pi \vartheta(b_x)  \label{Spherical-coord-varphi}  \\
&=&  \displaystyle \arctan\left( \frac{2  \mbox{\small sgn}(k)   \sin \phi_{SO}}{ b_x}\sqrt{\varepsilon_0(k)E_{SO,N}}\right)   \nonumber\\
& & \, +\pi \vartheta(b_x) \quad, 
\nonumber  
\end{eqnarray} 
where  the identities $\alpha_N^2 k^2=4   \varepsilon_0(k)\, E_{SO,N}$ and $b_x^2=E_Z^2$ have been used. 
By inverting the spectral relations (\ref{spectrum-N-side-e-branch}) and (\ref{spectrum-N-side-h-branch}) for the two bands $E_{N,e/h,\pm}(k)=E$ in favor of $k$, one obtains the set  of wavevectors $k = k(E)$ at  given energy $E$ in the N-side, which always come in pairs $\left(k(E), -k(E)\right)$  since $E_{N,e/h,\pm}(k)$ are even in k.
As a consequence, if the generic solution $k(E)$ represents an incoming mode, i.e. $k(E) = k^a(E)$, it follows that $-k(E)$ represents an outgoing mode, i.e. $-k(E) =  -k^a(E) = k^b(E)$, and viceversa. Thus, in the case of propagating modes, for each pair  $\left(k(E), -k(E)\right)$, we shall provide only the incoming mode solution, labeled by \textit{a}, while for the evanescent modes we shall write only the physically acceptable in the N-side, i.e. the   the solution with negative imaginary part.  

For each band ($b=\pm$) and each branch ($p=e/h$), the number and the type (propagating/evanescent)  of $k$-solutions of the equation $E_{N,e/h,\pm}(k)=E$  strictly depend on the specific values of the  parameters $\mu_N\,, E_{SO,N}, \, E_Z$. In particular, depending on the values of $E_{SO,N}$ and $E_Z$, one can distinguish  three different regimes, namely the Rashba dominated regime ($0<E_Z<2E_{SO,N}$), the weak Zeeman regime ($2E_{SO,N}<E_Z<4E_{SO,N}$) and the strong Zeeman regime ($4E_{SO,N}<E_Z$).
 Moreover, within each of these regimes, the chemical potential $\mu_N$ further determines various  ranges for the energy $E$, in which the $k$-wavevectors solving $E_{N,e/h,\pm}(k)=E$ take different expressions.
Here below, we shall explicitly provide such expressions in each of the three aforementioned regimes, in terms of the following quantities
\begin{align}
        A(E) =& \frac{2m^*}{\hbar^2}\left(\mu_N+E+2E_{SO,N}\right) \label{A(E)-def}\\
        B(E) =& \frac{2m^*}{\hbar^2}\sqrt{4E_{SO,N}\left(\mu_N+E+E_{SO,N}  \right) + E_Z^2} \label{B(E)-def} \\
\label{B-prime}
    {C}(E) =& \frac{2m^*}{\hbar^2}\sqrt{4E_{SO,N}\left(E-\mu_N-E_{SO,N}\right) -E_Z^2}\quad.
\end{align}

\subsubsection{Rashba dominated regime ($0< E_Z< 2 E_{SO,N}$)}
Within this regime, the value $\mu_N$ of the chemical potential  identifies 
various energy ranges. For definiteness, we shall focus on the case 
 \begin{equation} \label{muN-range-considered}
     0\le\mu_N < \frac{1}{2}\left(E_Z -  \frac{E_Z^2}{4E_{SO,N}}\right)  \quad,
 \end{equation} 
which is the regime that we have considered for the plots shown in the Main Text. Thus, five energy ranges can be distinguished. \\

\begin{notainterna}
\leo{Il valore $I_2$ corrisponde all'energia $E_4$ del nostro regime topologico (Fig.2) ed al valore $E_2$ nel nostro regime triviale (Fig.3). Non so quanto sia effettivamente importante specificarlo, anche perchè nei nostri due casi (trivial e topological) abbiamo assunto $\mu_N = 0$, quindi in realtà sià l'energia $E_4$ nel topological che l'energia $E_2$ trivial individuano sia il valore $E_Z -\mu_N$ che il valore $E_Z +\mu_N$.}
\end{notainterna}

\noindent {\it Range 1:} $0\le E <\mu_N + \frac{E_Z^2}{4E_{SO,N}}$
    \begin{equation}
        \begin{aligned}
            k^c_{N,e,+} &=  -i \sqrt{B(E) - A(E)}  \\[5pt]
        k^c_{N,h,+} &=  -i \sqrt{B(-E) - A(-E)} \\[5pt]
        k^a_{N,e,-} &=  \sqrt{A(E) + B(E)}\\[5pt]
        k^a_{N,h,-,1} &=  -\sqrt{A(-E) + B(-E)}
        \end{aligned}\quad,
    \end{equation}
    where we assign the index ``1" to the $k^{a}_{N,h,-,1}$ wavevectors, since in the ranges 4 e 5 (see below), the $E_{N,h,-}$ branch will admit more than two solutions.
    For the incoming modes $ k^a_{N,e,-}\,,\;k^a_{N,h,-,1}$ the group velocities read
\begin{align}
    v_{N,e,-}(E)
&=\frac{\hbar}{m^*}\, k^{a}_{N,e,-}(E)\;
\frac{B(E)}{\,4\frac{m^*}{\hbar^2}E_{SO,N}+B(E)} \label{velocities-Electron-banda-minus} \\\nonumber\\
v_{N,h,-,1}(E)
&=-\frac{\hbar}{m^*}\,k^{a}_{N,h,-,1}(E)\;
\frac{B(-E)}{\,4\frac{m^*}{\hbar^2}E_{SO,N}+B(-E)}  \label{velocities-Hole-banda-minus-1} 
\end{align}

\noindent {\it Range 2:} $\mu_N + \frac{E_Z^2}{4E_{SO,N}} \le E < E_Z-\mu_N$ 
\begin{equation}
    \begin{aligned}
       k^c_{N,e,+} &=  -i \sqrt{B(E) - A(E)} \\[5pt]
        k^a_{N,e,-} &=  \sqrt{A(E) + B(E)} \\[5pt]
        k^a_{N,h,-,1} &=  -\sqrt{A(-E) + B(-E)}\\[5pt]
        k^c_{N,h,-,1} &= - i \sqrt{B(-E) - A(-E)}
    \end{aligned}
\end{equation}
    with the group velocities magnitude of the two incoming modes already computed in Eqs. \eqref{velocities-Electron-banda-minus}-\eqref{velocities-Hole-banda-minus-1}.\\

\noindent {\it Range 3:} $E_Z-\mu_N \le E < E_Z+\mu_N$
\begin{equation}
    \begin{aligned}
         k^a_{N,e,+} &=  \sqrt{A(E) - B(E)}  \\[5pt]
        k^a_{N,e,-} &=  \sqrt{A(E) + B(E)} \\[5pt]
        k^a_{N,h,-,1} &= -  \sqrt{A(-E) + B(-E)} \\[5pt]
        k^c_{N,h,-,1} &=  -i \sqrt{B(-E) - A(-E)}
    \end{aligned}
\end{equation}
where the group velocity for $k^{a}_{N,e,+}$ mode is
\begin{align} \label{velocities-Electron-banda-plus}
    v_{N,e,+}(E)
&=\frac{\hbar}{m^*}\,k^{a}_{N,e,+}(E)\;
\frac{B(E)}{B(E)-4\frac{m^*}{\hbar^2}E_{SO,N}}
\end{align}

\noindent {\it Range 4:}  $E_Z+\mu_N \le E \le E_{SO,N}+ \mu_N+\frac{E_Z^2}{4E_{SO,N}}$ 
\begin{equation}\label{N-Rashba-dom-range-4}
    \begin{aligned}
         k^a_{N,e,+} &=  \sqrt{A(E) - B(E)} \\[5pt]
        k^a_{N,e,-} &=  \sqrt{A(E) + B(E)} \\[5pt]
         k^a_{N,h,-,1} &=  -\sqrt{A(-E) + B(-E)} \\[5pt]
         k^a_{N,h,-,2} &=   \sqrt{A(-E) - B(-E)}
    \end{aligned} \quad,
\end{equation}
with the group velocity for $k^a_{N,h,-,2}$ incoming mode being
\begin{align} \label{velocities-Hole-banda-minus-2}
   v_{N,h,-,2}(E)
&=\frac{\hbar}{m^*}\,k^a_{N,h,-,2}(E)\;
\frac{B(-E)}{\,4\frac{m^*}{\hbar^2}E_{SO,N}-B(-E)} \quad.
\end{align}

\noindent {\it Range 5:} $E > E_{SO,N}+ \mu_N+\frac{E_Z^2}{4E_{SO,N}}$
\begin{equation} \label{Rashba-dominated-Range5}
    \begin{aligned}
        k^a_{N,e,+} &=  \sqrt{A(E) - B(E)} \\[5pt]
        k^a_{N,e,-} &=  \sqrt{A(E) + B(E)} \\[5pt]
        k^c_{N,h,-,1} &=  -\sqrt{A(-E) +i {C}(E)} \\[5pt]
         k^c_{N,h,-,2} &=   \sqrt{A(-E) - i {C}(E)}  
    \end{aligned}
\end{equation}
where ${C}(E)$ is defined in Eq.(\ref{B-prime}).  
In order to highlight the negative imaginary part, it is possible to rewrite $k^c_{N,h,-,1}$ and $k^c_{N,h,-,2}$ as
\begin{align}
    k^c_{N,h,-,1} =  -\sqrt{Z(E)} \quad \qquad
    k^c_{N,h,-,2} =  +\sqrt{Z^*(E)}
\end{align}
with 
\begin{align}
    \sqrt{Z(E)} &=\nonumber \sqrt{\frac{|Z(E)|+ A(-E)}{2}} +i \sqrt{\frac{|Z(E)|- A(-E)}{2}} \\[5pt]
    |Z(E)| &= \sqrt{[A(-E)]^2 + [C(E)]^2}\quad.
\end{align} 
Note that two  energy threshold values $I_1 = \mu_N +  E_Z^2/4E_{SO,N}$ and $I_2 = E_Z - \mu_N$ appear in the above energy ranges. The assumption (\ref{muN-range-considered}) implies the relation $0<I_1 < I_2$. For   chemical potential values outside the range (\ref{muN-range-considered}), such relation does not hold. Then, the various energy ranges and the expressions of the wavevectors slightly change, although they can be obtained similarly.

 \subsubsection{Weak Zeeman regime ($2 E_{SO,N} \le E_Z < 4E_{SO,N}$)}
By still assuming that the chemical potential $\mu_N$ fulfills the condition (\ref{muN-range-considered}), the weak Zeeman regime identifies the very same five energy ranges as the Rashba dominated regime, where it exhibits the  same expressions of the wavevectors,  except for  the energy range~4, where Eqs.(\ref{N-Rashba-dom-range-4}) should be now replaced with:

\noindent {\it Range 4:}  $E_Z+\mu_N \le E \le E_{SO,N}+\frac{E_Z^2}{4E_{SO,N}}+ \mu_N $.  
\begin{equation}
    \begin{aligned}
        k^a_{N,e,+} &=  \sqrt{A(E) - B(E)}  \\[5pt]
        k^a_{N,e,-} &=  \sqrt{A(E) + B(E)} \\[5pt]
         k^c_{N,h,-,1} &= -i \sqrt{-A(-E) - B(-E)} \\[5pt]
         k^c_{N,h,-,2} &= -i \sqrt{B(-E) - A(-E)}
    \end{aligned}
\end{equation}
The group velocities of the (incoming) propagating modes $ k^a_{N,e,+}\;\text{and}\;\, k^a_{N,e,-}$ are given, respectively, in Eq.\eqref{velocities-Electron-banda-plus} and Eq.\eqref{velocities-Electron-banda-minus}.

\subsubsection{Strong Zeeman regime ($E_Z > 4 E_{SO,N}$)}
Also in this regime, the value of the chemical potential determines various energy ranges for the expressions of the wavevectors. Here, we shall assume that the condition
\begin{equation}
    0 \le \mu_N< E_Z  
\end{equation}
is fulfilled. Then, one has the following five ranges.\\

\noindent {\it Range 1:}  $0 \le E < E_Z-\mu_N$
    \begin{equation}
        \begin{aligned}
        k^c_{N,e,+} &=  -i \sqrt{B(E) - A(E)}\\[5pt]
        k^c_{N,h,+} &= - i \sqrt{B(-E) - A(-E)}\\[5pt]
        k^a_{N,e,-} &=  \sqrt{A(E) + B(E)}\\[5pt]
        k^a_{N,h,-} &= -  \sqrt{A(-E) + B(-E)}
        \end{aligned}
    \end{equation}
\noindent {\it Range 2:}  $E_Z - \mu_N \le E \le E_Z+\mu_N$
    \begin{equation}
        \begin{aligned}
        k^a_{N,e,+} &=  \sqrt{A(E) - B(E)}\\[5pt]
        k^c_{N,h,+} &= - i \sqrt{B(-E) - A(-E)}\\[5pt]
        k^a_{N,e,-} &=  \sqrt{A(E) + B(E)}\\[5pt]
        k^a_{N,h,-} &= -  \sqrt{A(-E) + B(-E)}
        \end{aligned}
    \end{equation}
\noindent {\it Range 3:} $E_Z + \mu_N < E < \mu_N + \frac{E_Z^2}{4 E_{SO,N}}$
    \begin{equation}
        \begin{aligned}
        k^a_{N,e,+} &=  \sqrt{A(E) - B(E)}\\[5pt]
        k^c_{N,h,+} &= - i \sqrt{B(-E) - A(-E)}\\[5pt]
        k^a_{N,e,-} &=  \sqrt{A(E) + B(E)}\\[5pt]
        k^c_{N,h,-,1} &= -i \sqrt{-A(-E) -B(-E)}
        \end{aligned}
    \end{equation}
\noindent {\it Range 4:} $\mu_N + \frac{E_Z^2}{4 E_{SO,N}} < E \le E_{SO,N} + \mu_N + \frac{E_Z^2}{4 E_{SO,N}}$
     \begin{equation}
        \begin{aligned}
            k^a_{N,e,+} &=   \sqrt{A(E) - B(E)}\\[5pt]
        k^a_{N,e,-} &=  \sqrt{A(E) + B(E)}\\[5pt]
        k^c_{N,h,-,1} &= -i \sqrt{-A(-E) -B(-E)}\\[5pt]
        k^c_{N,h,-,2} &= - i \sqrt{B(-E) - A(-E)}
        \end{aligned}
    \end{equation}
\noindent {\it Range 5:}  $E > E_{SO,N} + \mu_N + \frac{E_Z^2}{4 E_{SO,N}}$
   \begin{equation} 
    \begin{aligned}
        k^a_{N,e,+} &=  \sqrt{A(E) - B(E)} \\[5pt]
        k^a_{N,e,-} &=  \sqrt{A(E) + B(E)} \\[5pt]
        k^c_{N,h,-,1} &=  -\sqrt{A(-E) +i {C}(E)} \\[5pt]
         k^c_{N,h,-,2} &= \sqrt{A(-E) - i {C}(E)}    
    \end{aligned}
\end{equation}
where ${C}(E)$ has already been defined in \eqref{B-prime}.
The group velocities of the propagating modes have already been defined in Eq.\eqref{velocities-Electron-banda-plus},\eqref{velocities-Hole-banda-minus-1},\eqref{velocities-Hole-banda-minus-2},\eqref{velocities-Electron-banda-minus}.

\subsection{Solutions in the S-side}
In the bulk of the proximitized side S, the Hamiltonian is again  diagonal in $k$-space, $
\mathcal{H}=(1/2) \sum_k \Psi^\dagger_k H_{BdG}(k) \Psi_k^{}\,\,\,
$, where, however, the BdG Hamiltonian now contains   a finite superconducting pairing $\Delta_0 \neq 0$ and the direction of the SO field (\ref{alphaN-S-vec-def}) differs from the N-side. Explicitly, 
\begin{eqnarray} 
\lefteqn{H_{BdG}(k) = } & & \label{HBdG(k)-S}\\
&=& \begin{pmatrix}
     \xi_S(k) \sigma_0  +\mathbf{m}_S(k)\cdot \boldsymbol{\sigma}    &   \Delta_0 e^{i \varphi} \sigma_0
     \\ \Delta_0 e^{-i \varphi} \sigma_0
     &  -\xi_S(k) \sigma_0 + \mathbf{m}_S(-k)\cdot \boldsymbol{\sigma} 
\end{pmatrix} \quad, \nonumber
\end{eqnarray}  
where $\xi_S(k) =\hbar^2 k^2/2 m^* -\mu_S$ and 
\begin{eqnarray}\label{m(k)-S-def}
\mathbf{m}_S(k) = - (  \boldsymbol{\alpha}_S k + \mathbf{b}) =\left(-b_x\, , 0  ,\, -\alpha_S k \, \right)\nonumber\quad.
\end{eqnarray}
The spectrum of the Hamiltonian Eq.(\ref{HBdG(k)-S}) is given by Eq.(\ref{spectrum-S-side}), whereas its eigenstates are given by 
\begin{eqnarray}\label{w_S-pm}
\lefteqn{\mathsf{w}_{S,\pm}(k) =\frac{\mathcal{N}_k}{\Delta_0\left(b_x^2 +\alpha_S k \xi_S(k) \pm \Gamma(k)\right)} \,\,\times  } & & \\
& & \nonumber \\
& & \hspace{-0.5cm} \begin{pmatrix}
    - b_x e^{i\varphi}\left[  \xi^2_S(k) + \Delta_0^2 \, \pm \left( \Gamma(k)+\xi_S(k)E_{S,\pm}(k)\right)\right]\\ \\
    \begin{array}{l}  e^{i\varphi}\left[ b_x^2\xi_S(k)+( \alpha_S k\, \xi_S(k)\pm \Gamma (k)) \times \right. \\
    \hspace{2cm} \left. (\alpha_Sk+\xi_S(k)+E_{S,\pm}(k))\right]\!\!\end{array}\\ \\
    -b_x\Delta_0\left(E_{S,\pm}(k) + \alpha_S k\right)\\ \\
    \Delta_0\left(b_x^2 +\alpha_S k \,\xi_S(k) \pm \Gamma(k)\right)
 \end{pmatrix} , \nonumber 
\end{eqnarray}
where  
\begin{equation}
    \Gamma(k) = \sqrt{\xi^2_S(k) (\alpha_S^2 k^2 + b_x^2) +b_x^2\Delta_0^2} \quad,
\end{equation} 
and $\mathcal{N}_k $ is a normalization constant.
The electron-like or hole-like character of the   eigenstates (\ref{w_S-pm}) at a given $k(E)$ can be attributed through the sign of the quantity $\mathcal{Q}_\pm(k)$ defined in Eq.(\ref{charge character}).\\
 
\noindent The {\it electron-like branch} corresponds to $\mbox{sgn}(\mathcal{Q}_\pm(k))=+1$, and therefore has a spectrum
\begin{equation}\label{S-side-spectral-e-branch}
        E_{S,e,\pm}(k)  = \vartheta(+\mathcal{Q}_\pm(k)) \,E_{S,\pm}(k) \quad,
\end{equation}
where $E_{S,\pm}(k)$ are defined in Eq.\eqref{spectrum-S-side}, and corresponding eigenvector
\begin{equation}
        \mathsf{w}_{S,e,\pm}(k)  = \vartheta(+\mathcal{Q}_\pm(k)) \,\mathsf{w}_{S,\pm}(k)    \quad.
\end{equation}
The {\it hole-like branch}, corresponding to $\mbox{sgn}(\mathcal{Q}_\pm(k))=-1$,   has a spectrum
\begin{equation}\label{S-side-spectral-h-branch}
E_{S,h,\pm}(k)   = \vartheta(-\mathcal{Q}_\pm(k)) \,E_{S,\pm}(k) 
\quad,
\end{equation}
with eigenvectors
\begin{equation}
\mathsf{w}_{S,h,\pm}(k)  = \vartheta(-\mathcal{Q}_\pm(k)) \,\mathsf{w}_{S,\pm}(k) 
\quad.
\end{equation}
Similarly to the approach used for the N-side of the junction, here the wavevectors can be obtained by inverting the spectral relations \eqref{S-side-spectral-e-branch} and \eqref{S-side-spectral-h-branch} in favor of $k$, i.e. by solving $E_{S,e/h,\pm}(k)=E$. However, differently from the N-side, no analytical approach is viable to obtain the wavevectors $k = k(E)$ for the S-side. Thus, they have been derived via numerical solution, for each energy $E$.
\section{Details about the solution of the Scattering Matrix problem}
\label{AppB}
In this Appendix, we provide more technical details about the solution of the Scattering problem outlined in Sec.\ref{sec3a}. At each  given energy $E$, and   in each side $L=N,S$ of the interface, one has a  number $N_{pr,L}$   of  propagating(pr) channels, where each channel   exhibits a    mode  incoming  towards the interface and a mode  outgoing  from the interface.  Thus, in each side, one has $2N_{pr,L}$ unknown amplitudes for propagating modes. Furthermore, each side can also exhibit    conjugate pairs of complex modes. However, as mentioned above,   only one pair partner describes a  physically acceptable mode that does not diverge exponentially, i.e. an evanescent(ev) mode. Let $N_{ev,L}$ denote the number of evanescent modes in side $L=N,S$ at the given energy $E$. Since the equation $E_L(k)=E$ is ultimately a polynomial equation of 8th degree [see Eqs.(\ref{spectrum-N-side})-(\ref{spectrum-S-side})],   there are always 8 solutions in each side. Thus, the following relation holds
\begin{equation}\label{unknowns-1}
2(N_{pr,L}+N_{ev,L})=8   \qquad L=N,S\quad, 
\end{equation}
whereas the number of physical mode amplitudes  is  
\begin{equation}\label{unknowns-2}
N_L = 2N_{pr,L}+N_{ev,L}  \quad.
\end{equation}
Recalling that incoming and outgoing mode amplitudes are labelled by $a$ and $b$, respectively, while the   evanescent mode amplitudes are labelled by $c$, one can denote by 
\begin{equation}
\mathbf{x}_L = \left( \underbrace{a \ldots b \ldots}_{2N_{pr;L}}, \underbrace{c \ldots}_{N_{ev;L}} \right)^T \,   \qquad L=N,S\quad,
\end{equation}
the vector of the $N_L$ unknown amplitudes on  the side $L=N,S$. 
Then, the boundary conditions Eqs.(\ref{bc-fun})-(\ref{bc-der}) can compactly be written as
\begin{equation}\label{bc-M}
 \mathsf{M}_N(x_0)  \,
 \mathbf{x}_N  = \mathsf{M}_S(x_0)  \,\mathbf{x}_S   \quad,
\end{equation}
where the $8 \times N_N$ matrix $\mathsf{M}_N(x_0)$ and the $8 \times N_S$ matrix $\mathsf{M}_S(x_0)$ contain the various waves (propagating or evanescent) in the N- and S-side, respectively, evaluated at $x_0$. 

As a whole, Eq.(\ref{bc-M}) is a set of 8 linear equations for $N_N+N_S$ unknown amplitudes. This implies that one can fix only 8 unknowns  in terms of the other  variables, which remain undetermined.  Exploiting Eqs.(\ref{unknowns-1}) and (\ref{unknowns-2}), it is straightforward to see that
total number of undetermined unknowns   \begin{equation}\label{unknowns-3}
N_N+N_S-8 =
   N_{pr,N}+N_{pr,S} =N_{pr}   
\end{equation}
coincides with the total number of propagating channels. 
Notice also that total number of  unknowns that can be determined is 
$8  = N_{pr,N}+N_{pr,S}+N_{ev,N}+N_{ev,S}$.
We shall fix the amplitudes $a$ of the   incoming  propagating modes as the undetermined unknowns. Thus, the boundary conditions (\ref{bc-M})   enable one to express all the $b$ amplitudes of the  outgoing  propagating modes and all the $c$ amplitudes of the evanescent modes in terms of the $a$ amplitudes of the incoming propagating modes, as follows
\begin{equation}\label{Sigma-def}
\underbrace{\begin{pmatrix}
    b \\ \vdots \\ \hline c
\end{pmatrix}}_{8 \times 1}=\underbrace{\begin{pmatrix}
    & & \\
    & S & \\ 
        & & \\\hline
            & & \\
\end{pmatrix}}_{\Sigma \,\,(8 \times N_{pr})} \,\,\underbrace{\begin{pmatrix}
a  \\ \\
\end{pmatrix}}_{N_{pr} \times 1} \quad,
\end{equation}
where the explicit expression of the $8 \times N_{pr}$  matrix $\Sigma$  can be obtained from numerically solving the linear system (\ref{bc-M}), at each energy value $E$.
In particular, the upper block $S$ of   $\Sigma$ is  a $N_{pr} \times N_{pr}$ matrix connecting the outgoing amplitudes to the incoming amplitudes, and is the  Scattering matrix appearing in Eq.(\ref{S-matrix}).

\bibliography{biblio}
\end{document}